\documentclass[11pt, notitlepage, tightenlines,aps, prd, reprint,superscriptaddress,nofootinbib]{revtex4-1}
\usepackage[utf8]{inputenc}

\usepackage{placeins}
\usepackage{mathtools}
\usepackage{amsfonts}
\usepackage{amssymb,amsmath}
\usepackage{color}
\usepackage[dvipsnames]{xcolor}
\definecolor{lcolor}{rgb}{0.5,0,0}
\definecolor{citcolor}{rgb}{0,0.3,0.0}
\usepackage[breaklinks,colorlinks,urlcolor=blue,citecolor=citcolor,linkcolor=lcolor]{hyperref}
\usepackage{graphicx}
\usepackage{subfigure}
\graphicspath{ {image/} }
\usepackage{multirow}

\newcommand*\diff{\mathop{}\!\mathrm{d}}
\usepackage{braket}
\usepackage{bm}
\usepackage{txfonts}
\usepackage{mathrsfs} % in order to use mathscr
\usepackage{slashed}
\usepackage{accents}
\usepackage{hepunits}
\usepackage[normalem]{ulem}

\usepackage{soul} %strike out some texts  \st{something}
\usepackage{cancel} %cross out some texts \cancel{sth} \xcancel{sth}

\usepackage{cleveref}
\crefname{section}{Sec.}{Secs.}
\crefname{figure}{Fig.}{Figs.}
\crefname{appendix}{Appendix}{Appendices}
\crefname{equation}{Eq.}{Eqs.}
\crefname{table}{Table}{Tables}
\Crefname{section}{Section}{Sections}
\Crefname{figure}{Figure}{Figures}
\Crefname{appendix}{Appendix}{Appendices}
\Crefname{equation}{Equation}{Equations}
\Crefname{table}{Table}{Tables}

\usepackage{tikz}
\usetikzlibrary{arrows.meta,calc,decorations.pathmorphing,backgrounds}

\pgfdeclarelayer{background}
\pgfdeclarelayer{foreground}
\pgfsetlayers{background,main,foreground}

\newcommand{\expconfig}[1]{{\langle #1 \rangle_{event}}}
\newcommand{\tr}[1]{\mathrm{tr}\left[ #1 \right]}

\newcommand{\GeV}{{{\,}\textrm{GeV}}}
\newcommand{\fm}{{{\,}\textrm{fm}}}

\newcommand{\nc}{N_\mathrm{c}}

\begin{document}

\bibliographystyle{apsrev4-1}

\title{Kinetic and canonical momentum broadening in the Glasma
}

\thanks{Authors are listed in alphabetical order}

\author{Dana Avramescu}
\email{dana.d.avramescu@jyu.fi}
\affiliation{Department of Physics, P.O. Box 35, FI-40014 University of Jyv\"{a}skyl\"{a},
Finland}
\affiliation{
Helsinki Institute of Physics, P.O. Box 64, FI-00014 University of Helsinki,
Finland
}

\author{Carlos Lamas}
\email{carloslamas.rodriguez@usc.es}
\affiliation{Instituto Galego de Fisica de Altas Enerxias (IGFAE), Universidade de Santiago de Compostela, E-15782 Galicia, Spain}

\author{Tuomas Lappi}
\email{tuomas.v.v.lappi@jyu.fi}
\affiliation{Department of Physics, P.O. Box 35, FI-40014 University of Jyv\"{a}skyl\"{a},
Finland}
\affiliation{
Helsinki Institute of Physics, P.O. Box 64, FI-00014 University of Helsinki,
Finland
}

\author{Meijian Li}
\email{meijianli@hust.edu.cn}
\affiliation{Instituto Galego de Fisica de Altas Enerxias (IGFAE), Universidade de Santiago de Compostela, E-15782 Galicia, Spain}
\affiliation{School of Physics, Huazhong University of Science and Technology, Wuhan 430074, China}

\author{Carlos A. Salgado}
\email{carlos.salgado@usc.es}
\affiliation{Instituto Galego de Fisica de Altas Enerxias (IGFAE), Universidade de Santiago de Compostela, E-15782 Galicia, Spain}
\affiliation{Axencia Galega de Innovaci\'on (GAIN), Xunta de Galicia, Galicia, Spain}

\begin{abstract}
 
We lay the foundations for a quantum formalism describing the real-time evolution of particles in the Glasma phase of a heavy-ion collision, focusing on the implications of gauge invariance in the definition of the momentum of a particle in a classical background field. We first establish the correspondence between the classical Wong's equations and the Heisenberg equations of motion for a particle in a classical non-Abelian background field. Using this correspondence, we obtain equations of motion for both the kinetic momentum — the gauge invariant, physically measurable quantity — and the canonical momentum, which is conjugate to the coordinates in the Hamiltonian. In particular, the kinetic momentum broadening receives non-trivial contributions from the transverse field components, even in the eikonal limit. Finally, we demonstrate that imposing a transverse Coulomb gauge condition at the initial time significantly reduces the accumulation of numerical errors, thereby providing an optimized framework for the forthcoming quantum implementation.

\end{abstract}

\maketitle

\tableofcontents

\section{Introduction}

At weak coupling, the matter produced in the early stages of relativistic heavy-ion collisions can be described as Glasma fields \cite{Kovner:1995ja,Lappi:2006fp,Fukushima:2011nq,Gelis:2012ri,Albacete:2014fwa}, formulated using the Color Glass Condensate (CGC) framework \cite{Iancu:2000hn, Iancu:2003xm, Gelis:2010nm, Gelis:2012ri}. The Glasma consists of strong classical gluon fields far from equilibrium, corresponding to nonperturbatively large occupation numbers of gluonic states. Hard probes, such as jets and heavy quarks, are formed early in the collision. However, while  their properties in the Quark Gluon Plasma (QGP) phase have been studied extensively \cite{Cunqueiro:2021wls,vanHees:2005wb,Apolinario:2022vzg}, the 
interaction with the initial stage is often neglected. Only recently, there has been growing interest in quantifying the effect of the Glasma on the dynamics of hard probes.
Most existing studies rely on classical transport equations \cite{Wong:1970fu} to investigate heavy quarks \cite{Ruggieri:2018rzi,Sun:2019fud,Carrington:2020sww,Khowal:2021zoo,Ruggieri:2022kxv,Oliva:2024rex,Pooja:2024rnn,Avramescu:2024poa,Avramescu:2024xts,Backfried:2024rub} and jets \cite{Ipp:2020mjc,Ipp:2020nfu,Carrington:2021dvw,Carrington:2022bnv,Avramescu:2023qvv,Carrington:2026qlg} propagating through the Glasma. It is noted from these studies that hard probes experience significant momentum broadening during the Glasma phase. A promising approach to go beyond the classical approximation for the probe, is to describe the jet as a quantum state and evolve it in real time using the light-front Hamiltonian formalism. 
This method has been successfully applied to jet evolution within a classical color field described by local Gaussian distributions \cite{Li:2020uhl,Li:2021zaw, Li:2023jeh,Li:2025wzq} and implemented with quantum simulations \cite{Barata:2022wim,Barata:2023clv,Qian:2024gph}; such a formulation is applicable to both the cold nuclear matter in deep inelastic scattering, and the hot dense medium in the QGP stage in heavy-ion collisions.

Although momentum broadening and medium-induced radiation of partons propagating through the QGP phase can both be evaluated within well-established theoretical frameworks \cite{Romatschke:2006bb,Casalderrey-Solana:2007knd,Baier:2008js,DEramo:2010wup,Mrowczynski:2016etf,Hauksson:2021okc,Andres:2022ndd,Hauksson:2023tze,Kuzmin:2023hko}, studies of parton propagation in the Glasma have primarily focused on momentum broadening. The current ways to address gluon radiative processes rely on simplifying assumptions for the Glasma. Using the classical particle formalism, energy loss and collisions between particles were considered in non-Abelian plasmas and were shown to trigger plasma instabilities \cite{Dumitru:2005gp,Dumitru:2006pz,Dumitru:2007rp,Bodeker:2007fw}. Recently, jet quenching in a Glasma-like medium was studied by considering soft gluon radiation for jets in correlation domains consisting of constant longitudinal electric fields \cite{Barata:2024xwy}. Such estimates are similar to calculations of synchrotron-like gluon emission in the Glasma \cite{Aurenche:2012qk}. Beyond the Glasma, the effect of the pre-equilibrium stage on heavy quark \cite{Das:2017dsh,Singh:2025duj} and jet \cite{Zigic:2019sth,Andres:2019eus,Andres:2022bql,Adhya:2024nwx,Pablos:2025cli} observables such as the nuclear modification factor $R_{AA}$ or elliptic flow $v_2$, was investigated using numerous methods. However, these studies have not converged on an overall agreement on the magnitude of the pre-equilibrium effect compared to the QGP. Moreover, jet and heavy quark momentum broadening and energy loss were estimated using Effective Kinetic Theory (EKT) of QCD for the pre-equilibrium phase subsequent to the Glasma \cite{Boguslavski:2023fdm,Pandey:2023dzz,Du:2023izb,Boguslavski:2024ezg,Boguslavski:2024jwr,Boguslavski:2025ylx,Altenburger:2025iqa,Barata:2025agq,Barata:2025zku,Danhoni:2026gve} and the transport coefficients were shown to be compatible with the values reported in the Glasma. All these recent results contain important steps toward more complete calculations of momentum broadening and energy loss in the pre-equilibrium stages. Nevertheless, it still remains uncertain what is the relative contribution of the early stages compared to the QGP phase, and what are the underlying physical mechanisms present in the pre-equilibrium stages. 

The motivation of this work is twofold. First, we aim to clarify the distinction between kinetic and canonical momentum broadening, which has not been addressed in previous studies of parton propagation in the Glasma or broader scenarios of other background fields. Second, we want to lay  the foundation for a future quantum implementation, which will be part of first-principle systematic studies of radiative processes in the Glasma. Previous studies which rely on the classical transport equations in the Glasma fields \cite{Ruggieri:2018rzi,Carrington:2020sww,Ipp:2020mjc,Ipp:2020nfu,Carrington:2021dvw,Carrington:2022bnv,Avramescu:2023qvv} extract the kinetic momentum broadening, which is the gauge invariant momentum in the presence of a background gauge field. This is different from the canonical momentum broadening, evaluated from the momentum conjugated to the coordinates, and which is gauge dependent. The canonical momentum is the quantity that naturally appears in a quantum mechanical calculation that is needed to include gluon radiation. In typical perturbative scattering calculations, it is not necessary to take into account this difference, as one works with asymptotic states far away from the interaction region where the background field is zero. However, when performing the real-time evolution of a particle inside the medium, it is fundamental to take into account this difference. The canonical momentum contains a large gauge-dependent contribution from the background field, that must be taken into account to obtain the physical gauge invariant kinetic momentum. In principle, analytical calculations based on simplifying approximations, such as weak coupling or early-time expansions, could remove this difference by an appropriate gauge choice. However, to our knowledge, no existing analytical work on the Glasma has addressed the distinction between kinetic and canonical momentum broadening. In numerical simulations of the Glasma, which are performed in a fixed gauge, this difference is automatically present, yet the effect of gauge transformations on the magnitude of the canonical momentum broadening has not been quantified before.

The difference between kinetic and canonical transverse momentum broadening comes from the transverse component of the gauge field. The common lore in jet quenching calculations is that, in the eikonal limit, only the longitudinal component of the gauge field contributes to transverse momentum broadening, as it couples to the large $p^+$ component of the parton momentum. This is a standard result in jet quenching formalisms \cite{Casalderrey-Solana:2007knd} and applies equally in the eikonal propagation of the CGC shockwave picture \cite{Iancu:2003xm, Dumitru:2002qt}. This is indeed true for the canonical momentum. However, as we will show in this paper for the case of the Glasma, for the gauge invariant kinetic momentum inside the medium, the transverse component of the gauge potential needs to be taken into account already at leading order in the high energy of the particle. Therefore, there is a fundamental difference between calculations which use asymptotic states, measured far away from the interaction region, and the situation where one measures the evolution of the particle inside the background field. This difference can be understood in terms of the distinction between canonical and kinetic momentum. While we demonstrate this explicitly for the Glasma, the underlying argument relies only on the general definition of kinetic and canonical momentum and is therefore expected to hold for a generic external classical field. We note that a recent work~\cite{Kar:2026vzk} on DIS dijet production in a background field approach also gets a non-trivial contribution from the transverse fields at leading eikonal order. We here provide an explanation in terms of the difference between kinetic and canonical momentum broadening.

In this paper, we study the kinetic and canonical momentum broadening of quarks traversing the Glasma fields. We explicitly demonstrate the correspondence between the classical equations and the quantum Heisenberg equations for the quark evolution in a non-Abelian background field. Thus, we establish the relation between the classical approach and the full quantum calculation, carried out in the parallel work~\cite{Avramescu:2026qro}. In the numerical results in this paper, we treat the quark as a classical particle whose propagation is governed by Wong's equations. This enables us to quantify the distinction between kinetic and canonical momentum broadening for both static and eikonal quarks in the Glasma fields. We present two alternative methods for evaluating the decomposition of the kinetic momentum broadening: one based on directly tracking the force acting on the quark, and another based on decomposing the kinetic momentum into its canonical and gauge field contributions. While both methods serve as a mutual consistency check, the latter approach — which explicitly involves the gauge potential — bears a closer connection to the quantum formulation and thereby provides a natural bridge to the quantum implementation in~\cite{Avramescu:2026qro}. In addition, we quantify how gauge transformations of the Glasma fields affect the canonical momentum broadening, and discuss the relation between the gauge choice and the accumulation of numerical errors. For this purpose, we  impose a transverse Coulomb gauge condition, which minimizes the squared gauge potential. We show that this gauge fixing  significantly reduces numerical errors when evaluating gauge-dependent quantities — a crucial consideration for the quantum formulation. We also point the reader to Ref.~\cite{Gelis:2005pb} where the Coulomb gauge condition was imposed to identify the quark momentum in a solution of the Dirac equation in the Glasma background.

This paper is organized as follows. In \cref{sec:formalism}, we introduce the formalism for the Glasma fields and the classical particles, and derive the quantities relevant for computing the momentum broadening. In \cref{sec:results}, we present and discuss the numerical results. We conclude in \cref{sec:summary}. 

\section{Formalism}\label{sec:formalism}

In this section, we first introduce the formulation of the Glasma fields in the CGC effective theory. We then discuss the propagation of a classical particle in these fields. Lastly, we investigate how they are affected by gauge transformations.
 
\subsection{The Glasma fields} \label{sec:GlasmaFields}

To study the propagation of particles through the initial stage of a heavy-ion collision, one needs a realistic description of the matter produced immediately after the collision. In the weak coupling regime, this stage is characterized by highly occupied, out-of-equilibrium classical gluon fields known as
the Glasma~\cite{Kovner:1995ja,Lappi:2006fp,Fukushima:2011nq,Gelis:2012ri,Albacete:2014fwa}. These fields emerge from the collision of two ultrarelativistic nuclei within the CGC framework~\cite{Iancu:2000hn, Iancu:2003xm, Gelis:2010nm, Gelis:2012ri}. 

The CGC is an effective field theory based on the separation of scales between partons carrying a large fraction of the longitudinal momentum of the nucleus (large $x$) and those carrying a small fraction (small $x$). The large-$x$ degrees of freedom, often interpreted as valence partons, are treated as static color sources, while the small-$x$ gluons are described as classical gauge fields radiated by these sources. The use of a classical field description is justified by the large gluon occupation numbers that arise in the small-$x$ regime. 

The color fields obey the classical Yang-Mills equation of motion
\begin{align} \label{eq:Yang-Mills}
    [\mathcal{D}_\mu, \mathcal{F}^{\mu\nu}] = \mathcal{J}^\nu \, ,
\end{align}
where $\mathcal{D}_\mu=\partial_\mu+ig \mathcal{A}_\mu$ is the covariant derivative, $\mathcal{F}^{\mu\nu}=\partial^\mu \mathcal{A}^\nu -\partial^\nu \mathcal{A}^\mu+ig[\mathcal{A}^\mu, \mathcal{A}^\nu]$ is the field strength tensor, and $\mathcal{J}^\nu$ denotes the color current generated by the large-$x$ color charges. We note by $\mathcal{A}_\mu$ the SU($\nc $) matrix valued gauge fields $\mathcal{A}_\mu=\mathcal{A}_\mu^at^a$, where $t^a$ are the generators of SU($\nc $) in a given representation $R$, with $a$ from $1$ to $\nc ^2-1$, and similarly  the field strength tensor $\mathcal{F}_{\mu\nu}=\mathcal{F}_{\mu\nu}^a t^a$. For a nucleus moving along the light-cone axis $x^\pm = (t \pm z)/\sqrt{2}$, the color current is
\begin{align} \label{eq:ClassicCurrent}
    \mathcal{J}_{(1,2)}^\mu(x)=\delta^{\mu\pm}\,\delta(x^\mp)\,\rho_{(1,2)}(\vec{x}_\perp)\;,
\end{align}
where $\mathcal{J}_{(1)} $ corresponds to the right-moving nucleus ($+$ component) and $\mathcal{J}_{(2)}$ to the left-moving nucleus ($-$ component).
The current $\mathcal{J}_{(1)} $ [$\mathcal{J}_{(2)}$] is static in its light-cone time $x^+$ ($x^-$), localized in the longitudinal direction $x^-$ ($x^+$), and its transverse structure is determined by the color charge density $\rho_{(1)}$ [$\rho_{(2)}$].

The Yang-Mills equation  \cref{eq:Yang-Mills} for a single nucleus can straightforwardly be solved in the covariant gauge $\partial^\mu\mathcal{A}^{\,\mathrm{cov}}_{\mu, \,(1,2)}=0$. In this gauge, the only non-vanishing component of the gauge field is $\mathcal{A}_{\mathrm{cov}}^{+}$ (or $\mathcal{A}_{\mathrm{cov}}^{-}$) for nucleus $(1)$ moving along $x^+$ (or nucleus $(2)$ along $x^-$), which we denote as $\mathcal{A}^{\mathrm{cov}}_{(1,2)}$. The field equation simplifies to a two-dimensional Poisson equation $(\Delta_\perp-m_g^2) \mathcal{A}^{\,\mathrm{cov}}_{\,(1,2)}=-\rho_{(1,2)}$ which can be solved in momentum space as
\begin{equation}\label{eq:PoissionIRUV}
    \mathcal{A}^{\,\mathrm{cov}}_{\,(1,2)}(\vec{x}_\perp)=\int_0^{\Lambda} k_\perp \diff k_\perp \int \diff \theta_{k_\perp}\,\dfrac{\rho_{(1,2)}(\vec{k}_\perp)}{k_\perp^2+m_g^2}
    % \exp[-i\vec{k}_\perp\cdot\vec{x}_\perp]
    e^{-i\vec{k}_\perp\cdot\vec{x}_\perp}\,.
\end{equation}
Here $m_g$ and $\Lambda$ denote the infrared (IR) and ultraviolet (UV) regulators. 

For the fields before the collision, we use the MV model~\cite{McLerran:1993ni, McLerran:1993ka, McLerran:1994vd} which assumes that the distribution of color charges is stochastic (random gaussian) and local
\begin{subequations} \label{eq:MVModel}
    \begin{align}
        &\expconfig{ \rho^a(\vec{x}_\perp)}=0\,,\\
        &\expconfig{\rho^a(\vec{x}_\perp)\,\rho^b(\vec{y}_\perp)}=g^2\mu^2\delta^{ab}\delta^{(2)}(\vec{x}_\perp-\vec{y}_\perp)\,,
    \end{align}
\end{subequations}
where $g^2\mu$ represents the color charge density. Each Glasma event has a different initial transverse color charge distribution $\rho^a(\vec{x}_\perp)$, and Glasma event averages $\expconfig{\dots}$ are performed over these color charges. The MV model parameter $g^2\mu$ is the only physical scale. Typically, $g^2\mu$ is proportional to the saturation momentum $Q_s$~\cite{Lappi:2007ku}. The saturation scale $Q_s$ may be interpreted as the momentum scale at which the number of gluons is large enough for gluon recombination processes to compensate the generation of more gluons, due to $1 \to 2$ splittings, leading to the phenomenon of gluon saturation. 

\begin{figure}[tb!] 
    \centering
    \includegraphics[width=0.5\textwidth]{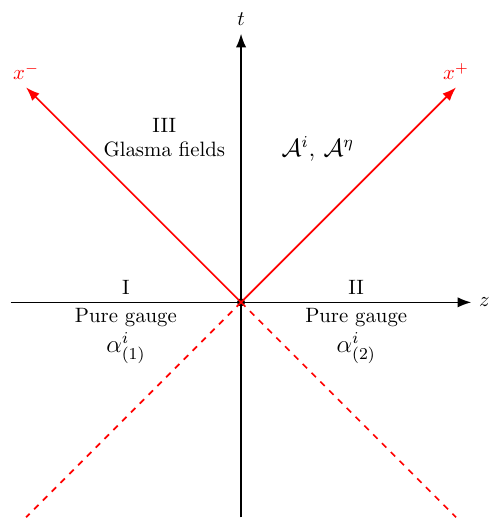}
    \caption{
    Illustration of the gauge potential in different regions: pure gauges from individual nucleus in Regions I and II, the resulting Glasma field after the collision in Region III, and vanishing fields in the backward light cone. 
    }
    \label{fig:LCDiagram}
\end{figure}

In order to obtain the Glasma fields, we transform the solutions for the single nucleus fields into the light-cone gauge $\mathcal{A}^{+,\mathrm{LC}}_{(1)} = 0$ ($\mathcal{A}^{-,\mathrm{LC}}_{(2)} = 0$) for a nucleus moving along $x^+$ ($x^-$). In this gauge, the nucleus moving in the $x^+$ ($x^-$) direction generates the gauge fields in Region I (II), as illustrated in \cref{fig:LCDiagram}. The fields in these regions reduce to transverse pure gauge fields
\begin{align} \label{eq:PureGaugeGCG}
    \alpha_{(1,2)}^{\underline{i}}(\vec{x}_\perp) = \frac{i}{g} \mathcal{V}_{(1,2)}(\vec{x}_\perp)\, \partial^{\underline{i}} \mathcal{V}_{(1,2)}^\dagger(\vec{x}_\perp)\; ,
\end{align}
with $\underline{i}\in(x,y)$ the transverse direction in the Glasma, and where we denote $\alpha_{(1,2)}^ {\underline{i}}\equiv\mathcal{A}^{\underline{i},\mathrm{LC}}_{(1,2)}$.
The gauge transformation is given by 
\begin{align}
    \mathcal{V}_{(1,2)}(\vec{x}_\perp) = \mathcal{P} \exp\left[ig\int \diff x^\mp \mathcal{A}^{\,\mathrm{cov}}_{\,(1,2)}(x^\mp, \vec{x}_\perp)\right]\,,
\end{align}
where $\mathcal{A}^{\,\mathrm{cov}}_{\,(1,2)}$ is the covariant gauge solution of  the Yang-Mills equation  \eqref{eq:Yang-Mills} found in Eq.~\eqref{eq:PoissionIRUV}.

To derive the initial conditions for the Glasma fields, we assume boost invariance, which is justified at sufficiently high collision energies. We express the Glasma fields in Region III in Milne coordinates, which are convenient in the description of boost-invariant expanding systems. The Milne coordinates consist of $\tau = \sqrt{t^2 - z^2}$ the proper time and $\eta = \tanh^{-1}(z/t)$ the space-time rapidity. Using the pure gauge fields in \cref{eq:PureGaugeGCG} before the collision and imposing boost-invariance $\partial_\eta\mathcal{A}^\mu=0$, one can derive the initial condition for the fields immediately after the collision in the Fock-Schwinger gauge $\mathcal{A}_\tau \equiv (x^+\mathcal{A}_+ + x^- \mathcal{A}_-)/\tau = 0$~\cite{Kovner:1995ts,Krasnitz:1998ns},
\begin{subequations} \label{eq:GlasmaInitialCondition}
    \begin{align} 
        &\mathcal{A}^{\underline{i}} (\tau = 0) = \alpha_{(1)}^{\underline{i}} + \alpha_{(2)}^{\underline{i}}\, , \\
        &\mathcal{A}^\eta (\tau = 0) = \frac{ig}{2}[\alpha_{(1)}^{\underline{i}}, \alpha_{(2)}^{\underline{i}}]\, ,
    \end{align}
\end{subequations}
along with $\partial_\tau\mathcal{A}^{\underline{i}} (\tau = 0)=0$ and $\partial_\tau\mathcal{A}^\eta (\tau = 0)=0$. Here, $\alpha_{(1,2)}^{\underline{i}}$ are the pure gauge transverse fields in Regions I ($x^->0, x^+<0$) and II ($x^+>0, x^-<0$) are known analytically from \cref{eq:PureGaugeGCG}, while $\mathcal{A}^\mu$ denotes the unknown Glasma fields in the future light-cone from Region III in~\cref{fig:LCDiagram}. The Fock-Schwinger gauge choice is convenient because it imposes that the field generated by one of the nuclei vanishes at the coordinate of the other nucleus (at $x^+ = 0$ it imposes $\mathcal{A}_- = 0$ and vice-versa). In this way, the fields do not interact with each other before the collision (there is no color precession of the color currents). 

Using the initial condition in \cref{eq:GlasmaInitialCondition}, the Glasma fields at later times can be obtained by solving the classical field equation of motion in the same Fock-Schwinger gauge. Since the static color charges are localized in $x^\pm$, initially there are no charges in Region III ($x^+>0$, $x^->0$). Consequently, the Glasma fields in this region evolve according to the sourceless Yang-Mills equation
\begin{align} \label{eq:Free_Yang-Mils}
    [\mathcal{D}_\mu, \mathcal{F}^{\mu\nu}] = 0\, .
\end{align}
The momenta conjugated to the fields are given by
\begin{subequations}  \label{eq:FieldMomenta}
    \begin{align}
        &\mathcal{P}^{\underline{i}}=\tau\,\partial_\tau\mathcal{A}_{\underline{i}}\,,\\
        &\mathcal{P}^\eta=\frac{1}{\tau}\,\partial_\tau\mathcal{A}_\eta\,,
    \end{align}
\end{subequations}
and the classical Yang-Mills equations in \cref{eq:Free_Yang-Mils} can be rewritten in terms of these momenta as 
\begin{subequations} \label{eq:CYMMomenta}
    \begin{align}
        &\partial_\tau \,\mathcal{P}^{\underline{i}} = \tau\, \left[ \mathcal{D}_{\underline{j}} ,\mathcal{F}_{\underline{j}\underline{i}}\right]
        -\frac{ig}{\tau}\, \left[\mathcal{A}_\eta,\left[\mathcal{D}_{\underline{i}},\mathcal{A}_\eta \right] \right]\,,\\
        &\partial_\tau \,\mathcal{P}^\eta = \frac{1}{\tau}\,
        \left[\mathcal{D}_{\underline{i}}, \left[\mathcal{D}_{\underline{i}},\mathcal{A}_\eta \right] \right]\,,
    \end{align}
\end{subequations}
where we used the boost-invariance $\partial_\eta \mathcal{A}_\mu=0$ and the temporal gauge condition $\mathcal{A}_\tau=0$.
There are several approximate semi-analytical approaches to solve the evolution equation, either by performing a proper time expansion valid for early times~\cite{Chen:2015wia,Carrington:2020ssh} or using the dilute approximation of the Glasma in which one solves linearized equations of motion~\cite{Kovner:1995ts,Krasnitz:1998ns, Blaizot:2008yb,Lappi:2017skr,Guerrero-Rodriguez:2021ask}. In this work, we use real-time lattice gauge theory~\cite{Krasnitz:1998ns} to numerically solve~\cref{eq:CYMMomenta} for the Glasma fields $\mathcal{A}_{\underline{i},\eta}$ and conjugate momenta $\mathcal{P}^{\,\underline{i},\eta}$. After discretizing the proper time $\tau$ as $\tau_n=n\Delta\tau$, the field equations from~\cref{eq:CYMMomenta} are solved using the leapfrog method, in which the fields are evaluated at integer time steps $\tau_n$ while the conjugate momenta at fractional time steps $\tau_{n+1/2}$.

The classical Yang-Mills equations in \cref{eq:CYMMomenta} are numerically  discretized using a transverse lattice of length $ L_\perp$ and $N_\perp$ lattice points in each direction, giving a lattice spacing $a_\perp=L_\perp/N_\perp$, along with periodic boundary conditions. To ensure gauge invariance of the discretized field equations, the evolution is expressed in terms of gauge link and plaquette variables. Due to boost invariance, the component $\mathcal{A}_\eta$ behaves effectively as an adjoint representation scalar field under the lattice discretization. The transverse gauge fields $\mathcal{A}_{\underline{i}}$ are used to construct gauge links
\begin{align} \label{eq:GaugeLink}
    U_{\underline{i}}\,(\tau,\vec{x}_\perp)= \exp\left[ig a_\perp \mathcal{A}_{\underline{i}}\left(\tau,\vec{x}_\perp+\frac{a_\perp}{2}\vec{e}_{\underline{i}}\right)\right]\,,
\end{align}
which is accurate up to $\mathcal{O}(a_\perp^2)$. Here, $\vec{e}_{\underline{i}}$ is the unit vector along a transverse direction $\underline{i}\in(x,y)$. The gauge links are Wilson lines connecting neighboring lattice sites, separated by the lattice spacing $a_\perp$. We denote a gauge link in the opposite direction to $\vec{e}_{\underline{i}}$ as $U_{-\underline{i}}\,(\vec{x}_\perp)=U_{\underline{i}}^\dagger(\vec{x}_\perp-\vec{a}_{\underline{i}})$, where we use the notation $\vec{x}_\perp \pm \vec{a}_{\underline{i}}$ for the transverse coordinate $\vec{x}_\perp \pm a_\perp\vec{e}_{\underline{i}}$.  The transverse gauge field may be approximately extracted as a logarithm of the corresponding gauge link
\begin{equation} \label{eq:GaugeField}
    \mathcal{A}_{\underline{i}}(\tau,\vec{x}_\perp)= -\frac{i}{ga_\perp}\mathrm{ln}\left[U_{\underline{i}}\,(\tau,\vec{x}_\perp)\right].
\end{equation}
On the lattice, the product of gauge links along the smallest lattice square defines a plaquette, namely $U_{\underline{i},\underline{j}}\,(\vec{x}_\perp)=U_{\underline{i}}\,(\vec{x}_\perp)\,U_{\underline{j}}\,(\vec{x}_\perp+\vec{a}_{\underline{i}})\,U_{-\underline{i}}\,(\vec{x}_\perp+\vec{a}_{\underline{i}}+\vec{a}_{\underline{j}})\,U_{-\underline{j}}\,(\vec{x}_\perp+\vec{a}_{\underline{j}})$ at a given $\tau$. The transverse field strength tensor $\mathcal{F}_{\underline{i}\underline{j}}$ corresponds to a plaquette variable
\begin{equation}
    U_{\underline{i},\underline{j}}\,(\tau,\vec{x}_\perp)=  \exp\left[ig a_\perp^2 \mathcal{F}_{\underline{i}\underline{j}}\left(\tau,\vec{x}_\perp+\frac{a_\perp}{2}\vec{e}_{\underline{i}}+\frac{a_\perp}{2}\vec{e}_{\underline{j}}\right)\right]\,,
\end{equation}
accurate up to $\mathcal{O}(a_\perp^3)$. Using this lattice discretization, we numerically solve \cref{eq:CYMMomenta} for the Glasma fields $\mathcal{A}_{\eta}$, the gauge links $U_{\underline{i}}$, and conjugate momenta $\mathcal{P}^{\,\underline{i},\eta}$ with the equation of motion for $\mathcal{P}^{\,\underline{i}}$ expressed using gauge links and plaquettes. 

The transverse $\underline{i}\in(x,y)$ and longitudinal $z$ color electric and magnetic fields in the Glasma can be expressed in terms of the Glasma gauge potentials along with the conjugate momenta from \cref{eq:FieldMomenta}
\begin{equation}\label{Eq:EBFieldsGlasma}
    \begin{aligned}
        &\mathcal{E}_{\underline{i}}  = \frac{1}{\tau}\,\mathcal{P}^{\underline{i}}=\partial_\tau\mathcal{A}_{\underline{i}}\,,
         &&\quad \mathcal{E}_z  = \mathcal{P}^\eta = \frac{1}{\tau\,}\partial_\tau \mathcal{A}_\eta\,, \\
        &\mathcal{B}_{\underline{i}}  = \frac{1}{\tau}\, \epsilon_{\underline{i}\underline{j}}\left[\mathcal{D}_{\underline{j}},\mathcal{A}_\eta\right],
         &&\quad \mathcal{B}_z  = -\mathcal{F}_{xy}\,.
    \end{aligned}
\end{equation}
These expressions follow from the general definitions of the electric and magnetic fields, $\mathcal{E}_{i}=\mathcal{F}_{ti}$ and $\mathcal{B}_{i}=-\epsilon_{ijk}\mathcal{F}_{jk}/2$
[with the 3-dimensional spatial index $i\in(x,y,z)$], by writing them in Milne coordinates and in the temporal gauge $\mathcal{A}_\tau = 0$, along with imposing boost-invariance $\partial_\eta\mathcal{A}_\mu=0$ and evaluating them at mid-rapidity $\eta = 0$ \cite{Fujii:2008dd}. On the lattice, we extract the covariant derivative as
$\left[\mathcal{D}_{\underline{i}},\mathcal{A}_\eta(\vec{x}_\perp)\right]=[U_{\underline{i}}(\vec{x}_\perp)\mathcal{A}_\eta(\vec{x}_\perp+a_\perp\vec{e}_{\underline{i}})U_{\underline{i}}^\dagger(\vec{x}_\perp)-U_{-{\underline{i}}}(\vec{x}_\perp)\mathcal{A}_\eta(\vec{x}_\perp-a_\perp\vec{e}_{\underline{i}})U_{-{\underline{i}}}^\dagger(\vec{x}_\perp)]/(2a_\perp)$ at a given $\tau$. The field strength is obtained from the plaquette as
$\mathcal{F}_{xy}=[U_{x,y}+U_{-x,y}+U_{x,-y}+U_{-x,-y}]_\mathrm{ah}/(4 a_\perp^2)$ at fixed $\tau$ and $\vec{x}_\perp$, where $[\dots]_\mathrm{ah}$ denotes the anti-Hermitian traceless part.

\subsection{Classical particles in Glasma}
\label{sec:ClassicaPropagation}

In this section, we describe the interaction of classical probes with the Glasma background field. We start by discussing Wong's equations, where the time dependence of the momentum arises from the non-Abelian generalization of the Lorentz force. We then present a derivation of Wong's equations as the classical limit of the quantum evolution of the corresponding operators. This method allows us to obtain the classical equations of motion for both the kinetic momentum, related to the motion of the particle, and the canonical momentum, conjugated to the particle coordinate in a Hamiltonian formalism. We also specialize these results to obtain expressions for the kinetic and canonical momentum broadening for both eikonal and static quarks. Lastly, we focus on the gauge dependence of the canonical momentum broadening and the decomposition of the particle Lorentz force. We explain the Coulomb gauge fixing procedure, which minimizes the gauge-dependent part. The results in this section are first obtained for a general non-Abelian gauge field. We then specialize our results to the case of  boost-invariant Glasma fields.

\subsubsection{Wong's equations} \label{sec:Wong}

\begin{table}[b!]
  \caption{\label{tab:A_convention} 
Conventions of the notations for the gauge field.
  }
  \centering
  \begin{ruledtabular}
  \begin{tabular}{lll}
    color matrix
    & $\mathcal{A}_\mu$
    & $\mathcal{A}_\mu^at^a$
    \\ 
         \hline
          classical color
    & $A_\mu$
    & $\mathcal{A}_\mu^a Q^a$\\
     \hline
          color operator
    & $\hat{A}_\mu$
    & $\mathcal{A}_\mu^a \hat{T}^a$ 
  \end{tabular}
  \end{ruledtabular}
\end{table}

The dynamics of a classical spinless point-like particle evolving in a classical Yang-Mills field $\mathcal{A}_\mu^a$ can be described by Wong's equations of motion~\cite{Wong:1970fu}
\begin{subequations} \label{eq:Wong}
    \begin{align}
        &\frac{\diff x^\mu}{\diff \lambda} = \frac{1}{m}p_{kin}^\mu(\lambda)\,,\\
        &\frac{\diff p^\mu_{kin}}{\diff \lambda}=\frac{g}{m} Q^a(\lambda) \,\mathcal{F}^{\mu\nu,a}(\lambda)\,p_\nu^{kin}(\lambda)\,,\\
        &\frac{\diff Q^a}{\diff \lambda}=-\frac{g}{m} f^{abc}Q^c(\lambda)\,\mathcal{A}_\mu^b(\lambda)\,p^\mu_{kin}(\lambda)\,,
    \end{align}
\end{subequations}
in which $x^\mu$ denotes the particle coordinate, $Q^a$ the classical color charge and $m$ is the particle mass. The kinetic momentum  $p^\mu_{kin}$ corresponds to the physical, measurable momentum of the particle, obtained as the derivative of the coordinate along the trajectory. Here $\lambda$ is the affine parameter used to parametrize the particle worldline, and the fields are evaluated along this particle trajectory. We can also write the equations in matrix form by introducing the color charge algebra elements $Q=Q^a t^a$. One can pass between the matrix and component notations using $Q^a X^a = \tr{QX}/T_R$ where $\mathrm{tr}[t^a t^b]=T_R\delta^{ab}$ and $T_R=1/2$ for the $R=F$ fundamental representation (quarks). It is is important to note that the generators $t^a$ are just a convenient notation to write equations in matrix form rather than with explicit color indices, and are distinct from quantum mechanical color charge operators that we will introduce later. For convenience, the different gauge-field notations used in this paper are summarized in \cref{tab:A_convention}. Using the matrix notation, one may express Wong's kinetic momentum and color charge evolution from~\cref{eq:Wong} as
\begin{subequations} \label{eq:WongAlg}
    \begin{align}
        &\frac{\diff p^\mu_{kin}}{\diff \lambda}=\frac{g}{T_R} \tr{Q(\lambda) \,\mathcal{F}^{\mu\nu}(\lambda)} \frac{\diff x_\nu}{\diff \lambda}\,,\\
        &\frac{\diff Q}{\diff \lambda}=ig [Q(\lambda),\mathcal{A}_\mu(\lambda)]\frac{\diff x^\mu}{\diff \lambda}\,.
    \end{align}
\end{subequations}
The color charge equation may be recast as a color rotation
\begin{align} \label{eq:WongColorRotation}
    Q(\lambda)=\mathcal{U}(\lambda;0)\,Q(0)\,\mathcal{U}^\dagger(\lambda;0)\,,
\end{align}
where the particle Wilson line
\begin{align} \label{eq:ParticleWilsonLine}
    \mathcal{U}(\lambda;0)=\mathcal{P}\exp\left[-ig\int_0^\lambda \diff\lambda^\prime \frac{\diff x^\mu}{\diff\lambda^\prime} \mathcal{A}_\mu(\lambda^\prime)\right]
\end{align}
picks up the background gauge potential along the particle trajectory. 

The classical color charges satisfy the quadratic Casimir invariant constraint 
\begin{align} \label{eq:ColorChargeCasimir}
    Q^aQ^a=q_2(R)\,.
\end{align}
Here, following \cite{Avramescu:2023qvv}, we normalize the color charges to satisfy $q_2(F)=C_2\nc = 4$ with  the ``classical Casimir invariant'' for fundamental representation quarks~\cite{Kelly:1994dh,Litim:1999id,Litim:2001db,Carrington:2016mhd,Ipp:2020mjc,Avramescu:2023qvv}, taken to be $\nc$ times the group theory Casimir $C_2=(\nc^2-1)/(2\nc)$. Additionally, the SU($3$) classical color charges satisfy the cubic Casimir constraint $q_3=d_{abc}Q^a Q^b Q^c$, with $q_3=\nc  C_3$, where $C_3=(\nc ^2-4)(\nc ^2-1)/(4\nc ^2)$ is the group theory Casimir for quarks. As discussed in Ref.~\cite{Avramescu:2023qvv},  it is in general not possible to construct explicit classical color charges with quadratic and cubic Casimirs corresponding to the fundamental representation of the quantum theory~\cite{Avramescu:2023qvv}. Thus we work in the classical theory with color charges that are different from the fundamental representation quantum theory, and then scale our results for the quark momentum broadening with the ratio of quadratic Casimir operators to obtain the expected color factor. The Casimir constraints are conserved by the evolution $\diff \,q_{2,3}/\diff\lambda=0$. 

The color charges are averaged over an ensemble of classical color charge configurations, where our choice of $q_2(F)=C_2\nc$ corresponds to 
\begin{align} \label{eq:ColorChargeIntegral}
    \int \diff Q \, Q^a Q^b= T_R \delta^{ab}\,,
\end{align}
along with the normalization $\int \diff Q =1$ and similarly $\int \diff Q\,Q^a =0$. We denote the classical color charge averages as 
\begin{equation}\label{eq:defintq}
    \langle X\rangle_Q\equiv\int \diff Q\,X\,,
\end{equation}
where $X$ is any $n$-point function of the classical color component $Q^a$. Note that we only focus on the quadratic Casimir from~\cref{eq:ColorChargeCasimir}, since the quantities we compute with this formalism, such as momentum broadening, will be mostly insensitive to the cubic Casimir of SU($3$)~\cite{Ipp:2020mjc,Avramescu:2023qvv}.

The classical momentum broadening along the $i\in(x,y,z)$ direction is defined as the square of the momentum variation 
\begin{align} \label{eq:ClassicBroadening}
    \delta p_i^{kin}(\lambda)=p_i^{kin}(\lambda)-p_i^{kin}(0)\,.
\end{align}
For the discussion that follows, it is useful to introduce a notation for
the color Lorentz force $f_i$ experienced by the particles while propagating in the background fields
\begin{equation}\label{eq:ClassicLorentzForce}
    f_i=\frac{\diff x^\mu}{\diff \lambda}\mathcal{F}_{i\mu}. 
\end{equation}
This Lorentz force has a color index, and to get a change in the particle momentum (which is a color singlet observable), this color index has to be contracted with the color charge. In terms of the color Lorentz force, the Wong's equation for the kinetic momentum  from \cref{eq:WongAlg} may equivalently be expressed  as
\begin{equation}\label{eq:KinMomBroadLorFor}
    \frac{\diff p_i^{kin}}{\diff \lambda}=\frac{g}{T_R}\tr{Q f_i}\,.
\end{equation}
We can now write an analytical expression for the momentum broadening resulting from Wong's equations for  the momentum~\cref{eq:WongAlg} and the color precession~\cref{eq:WongColorRotation} as
\begin{multline} \label{eq:MomBroadQaQb}
    (\delta p_i^{kin}(\lambda))^2=\frac{g^2}{T_R^2}Q^a(0)Q^b(0) \\
    \times \int_0^\lambda \diff \lambda^\prime\int_0^\lambda \diff\lambda^{\prime\prime} \tr{t^a \tilde{f}_i(\lambda^\prime)}\,\tr{t^b \tilde{f}_i(\lambda^{\prime\prime})}\,.
\end{multline}
Here the color Lorentz force is parallel transported with the particle Wilson line involved in the color charge rotation as
\begin{equation}\label{eq:ColorRotLorentzForce}
    \tilde{f}_i(\lambda)=\mathcal{U}(\lambda;0)\,f_i(\lambda)\,\mathcal{U}^\dagger(\lambda;0)\,.
\end{equation}
Note that this does not mean that we can solve Wong's equations analytically, since the expression involves a particle trajectory for which we do not have an explicit expression in the general case. 

Performing the classical color averages using the $2$-point functions from~\cref{eq:ColorChargeIntegral} in~\cref{eq:MomBroadQaQb} gives the broadening
\begin{align} \label{eq:KinMomLorentzForce}
    \langle (\delta p_i^{kin}(\lambda))^2\rangle_Q=g^2\int_0^\lambda \diff \lambda^\prime \int_0^\lambda \diff\lambda^{\prime \prime}\,\tr{ \tilde{f}_i(\lambda^\prime)\tilde{f}_i(\lambda^{\prime \prime})}\,,
\end{align}
where we used $\tr{t^a X}\,\tr{t^a Y}=T_R\,\tr{XY}$ and in which $\braket{\cdots}_Q$ denotes the color average defined in~\cref{eq:defintq}. As can be seen from \cref{eq:MomBroadQaQb}, the momentum broadening is proportional to the classical Casimir $q_2$. As discussed above, the charges in the classical calculation are normalized as $q_2= \nc C_2$ rather than with the fundamental representation Casimir $C_2$. Thus we follow here procedure also introduced in \cite{Ipp:2020mjc,Ipp:2020nfu,Avramescu:2023qvv} and perform calculations with the classical Casimir $q_2$, and then rescale our results for momentum  broadening with $C_2/q_2 = 1/\nc$, unless otherwise stated.

\subsubsection{Wong's equations in the Glasma}

Let us now specialize these results to the  Glasma  background field configuration, with details of its construction presented in~\cref{sec:GlasmaFields}. We focus on two limiting cases in which the quark trajectories are known and the kinetic momentum broadening in~\cref{eq:KinMomLorentzForce} reduces to correlators of Glasma electric and magnetic fields \cite{Avramescu:2023qvv}. These cases are an eikonal quark, corresponding to the $E\to\infty$ limit where the trajectory is a straight line, and a static quark in the $m\to\infty$ limit, where the trajectory is a fixed coordinate.

For an eikonal quark which propagates along the $x$-axis with $E\rightarrow\infty$, the trajectory reduces to $x(t)=t$, $y(t)=y(0)$, and $z(t)=z(0)$. We choose the affine parameter in~\cref{eq:ClassicLorentzForce} to be the coordinate time $\lambda = t$. Additionally, at mid-rapidity $\eta = 0$ the Glasma proper time is the same as the coordinate time $t = \tau$. Using that $E\, d/dt = m\, d/d\lambda$ the Wong's equation for transverse kinetic momentum can be expressed as
\begin{align}
    \frac{\diff p^i_{kin}}{\diff \tau} & = g Q^a(\tau) \left(\,\mathcal{F}^{it,a}(\tau) + \mathcal{F}^{i3,a}(\tau)\right)\, , 
\end{align}
where $i=y,z$. Therefore, an eikonal particle suffers a non-negligible change in its transverse momentum through the interaction with the medium, even when it is propagating along a straight line. A naive interpretation is that the longitudinal momentum of the particle is so large that the momentum broadening induced by the medium is not enough to significantly deflect the particle's trajectory. The Lorentz force experienced by the eikonal quark along the $i\in(y,z)$ direction reduces to
\begin{equation}\label{eq:LorentzForceTensorEik}
    f_i^{eik}(\tau)=\mathcal{F}_{i t}(\tau) + \mathcal{F}_{ix}(\tau)\;.
\end{equation}
This can be expressed in terms of the Glasma electric and magnetic field components~\cite{Ipp:2020mjc,Ipp:2020nfu,Avramescu:2023qvv}
\begin{subequations}\label{eq:JetLorentzForce}
    \begin{align} 
        &f^y_{eik}=\mathcal{E}_y-\mathcal{B}_z\,,\\ &f^z_{eik}=\mathcal{E}_z+\mathcal{B}_y\,,
    \end{align}
\end{subequations}
where the field components are expressed in terms of the Glasma gauge
fields in \cref{Eq:EBFieldsGlasma}. The proper time evolution of the transverse kinetic momentum components for eikonal quarks is given by~\cref{eq:KinMomBroadLorFor} upon substituting 
$f_i\to f_i^{eik} $. The color rotation according to~\cref{eq:ParticleWilsonLine} contains only the $x$-component of the Glasma gauge potential
\begin{align} \label{eq:JetColorRotation}
    \mathcal{U}_{eik}(\tau;0)=\mathcal{P}\exp\left[-ig\int_0^\tau \diff \tau^\prime \,\mathcal{A}_x(\tau^\prime)\right]\,.
\end{align}
Note that $\mathcal{A}_t$ vanishes due to the gauge choice and the evaluation at mid-rapidity. The classical kinetic momentum broadening averaged over quark color charge configurations is then given by \cref{eq:KinMomLorentzForce} upon taking $\lambda=\tau$ and substituting $\tilde{f}_i\to\tilde{f}^{eik}_{\,i}=\mathcal{U}_{eik}\,f^{eik}_i\mathcal{U}_{eik}^\dagger$.

Following a similar procedure, for a quark that has infinite mass $m\rightarrow\infty$, the trajectory is static, and the Wong's equations for the spatial components of the kinetic momentum read
\begin{align}
    \frac{\diff p^i_{kin}}{\diff t} = g Q^a(\tau) \,\mathcal{F}^{it,a} (\tau)\, , 
\end{align}
with $i=x,y,z$. Consequently, the heavy quark suffers a change in its momentum even if the trajectory is approximately static, as this increase in momentum is not enough to overcome the formally infinite quark mass. The Lorentz force for the static quarks is given only in terms of the electric fields~\cite{Avramescu:2023qvv}
\begin{equation}\label{eq:StatLorFor}
    f^i_{stat}=\mathcal{E}_i\,,
\end{equation}
with the Glasma electric fields expressed in~\cref{Eq:EBFieldsGlasma}. The $i\in(x,y,z)$ components of the static quark kinetic momentum are given by~\cref{eq:KinMomBroadLorFor} upon substituting $f_i\to f_i^{stat} $. In the temporal gauge $\mathcal{A}_\tau=0$, the color rotation from~\cref{eq:ParticleWilsonLine} has no effect, as $\mathcal{U}_{stat}=\mathcal{I}$. It follows that the classical kinetic momentum broadening is given by \cref{eq:KinMomLorentzForce} upon substituting $\tilde{f}_i\to\tilde{f}^{stat}_{\,i}=f^{stat}_i$, thereby receiving contributions from the electric fields only. This contrasts with the eikonal case, where both electric and magnetic fields contribute and the force is color rotated by the Wilson line along the quark trajectory.

\subsubsection{Quantum to classical correspondence} \label{sec:ClassicQuant}

The classical equations of motion can be obtained by taking the classical limit of the more general quantum formalism. 
To demonstrate the quantum-to-classical correspondence, we derive the Heisenberg equations of motion for the coordinate, kinetic momentum, color charge and spin operators of a quantum particle propagating in classical non-Abelian fields. Then, we show that these equations reduce to Wong's equations when taking the classical limit. For simplicity, we illustrate the quantum-to-classical correspondence for a particle with finite mass $m$, and specialize to the massless case only when discussing eikonal quarks. 

We start from the quadratic form of the Dirac Hamiltonian
\cite{Heinz:1984yq,Elze:1989un,Pooja:2022ojj} 
\begin{align} \label{eq:QuadDirac}
    \hat{H} = \frac{1}{2m}\left(m^2 - (\hat{p}^\mu - g \hat{A}^\mu)^2\right) - \frac{g}{2m} \hat{S}^{\mu\nu} \hat{F}_{\mu\nu} \, . 
\end{align}
This Hamiltonian can be obtained by squaring the Dirac operator, and thus it is clear that the solutions of the Dirac equation are its eigenstates. The zero eigenvalue reflects the mass-shell constraint $p^\mu p_\mu = m^2$, and corresponds to a line of constant phase of the wavefunction, as one would expect from the trajectory of a classical particle represented by a maximum in the wavefunction. We can therefore take it as the starting point for quantizing the theory, with $\lambda$ serving as a worldline parameter whose choice is a matter of convenience: $\lambda = \tau$ (proper time) preserves manifest Lorentz covariance, while $\lambda = t$ or $\lambda = x^+$ recover instant-form or light-front dynamics respectively. Here the minimal coupling $\partial_\mu \to \partial_\mu + ig\hat{A}_\mu$ encodes the interaction of the quantum particle with the classical background field. The quantum operator for coupling the particle to the gauge field $\hat{A}^\mu=\mathcal{A}^{\mu,a} \hat{T}^a$ is given in terms of the classical color component of the field $\mathcal{A}_\mu^a$ and the quantum color charge operator of the particle $\hat{T}^a$ and similarly for the field strength tensor $\hat{F}_{\mu\nu}=\mathcal{F}_{\mu\nu}^a\hat{T}^a$. Note that here the operators $\hat{T}^a$ are not the same as the purely numerical matrices $t^a$, which represent generators of the SU($\nc $) group. See also \cref{tab:A_convention} for a summary of the notation used in this work. The spin structure of the particle is contained in $\hat{S}^{\mu\nu} = \hat{\sigma}^{\mu\nu}/2$, where $\sigma^{\mu\nu} = i/2[\gamma^\mu, \gamma^\nu]$ is the antisymmetric tensor constructed from the Dirac gamma matrices.

To obtain the correspondence with the classical calculation, it is convenient to work in the Heisenberg picture, where all the information about time evolution is contained in the operators. In this picture, the operators satisfy the evolution equation
\begin{align}\label{eq:Ehrenfest}
    \frac{\diff \hat{\mathcal{O}}_H}{\diff \lambda} = i [\hat{H}_H, \hat{\mathcal{O}}_H] +  \frac{\partial \hat{\mathcal{O}}_H}{\partial \lambda}  \, ,
\end{align}
where $\lambda$ is the worldline proper time of the particle. For notation brevity, we omit the $H$ subscript below; all quantum operators are understood to be in the Heisenberg picture.

The velocity operator is the worldline proper time derivative of the coordinate operator 
\begin{align}\label{eq:VelocityOperator}
    \hat{v}^\mu (\lambda) \equiv \frac{d\hat{x}^\mu}{d\lambda} \, ,
\end{align}
which according to~\cref{eq:Ehrenfest} is given by
\begin{equation}
    \frac{d\hat{x}^\mu}{d\lambda} = i[\hat{H}, \hat{x}^\mu] = \frac{\hat{p}^\mu - g \hat{A^\mu}}{m}\;.
\end{equation}
Consequently, the kinetic momentum operator of the particle is given by
\begin{equation}\label{eq:KinCanMom}
    \hat{p}_{kin}^\mu \equiv m\frac{d\hat{x}^\mu}{d\lambda} = \hat{p}^\mu - g\hat{A}^\mu\,.
\end{equation}
In the presence of a background gauge field, the kinetic momentum differs from the canonical momentum $\hat{p}^\mu$, which is conjugated to the coordinates in the Hamiltonian. i.e. satisfies the canonical commutation relation. The equation of motion for the kinetic momentum may similarly be obtained using~\cref{eq:Ehrenfest} in the Heisenberg picture as
\begin{equation} \label{eq:HeisenbergKineticMomentum}
    \begin{split}
        \frac{\diff \hat{p}_{kin}^\mu}{\diff \lambda} & = i [\hat{H}, \hat{p}^\mu_{kin}] \\
        & = \frac{g}{2m}\left( \hat{p}_\nu^ {kin} \hat{F}^{\mu\nu} + \hat{F}^{\mu\nu} \hat{p}_\nu^{kin} - \hat{S}^{\nu\rho} \hat{D}^\mu \hat{F}_{\nu\rho}\right)\, ,
    \end{split}
\end{equation}
where $\hat{D}_\mu=\partial_\mu+ig \hat{A}_\mu$. Note that $\hat{A}^\mu$ and $\hat{F}^{\mu\nu}$ do not explicitly depend on the affine parameter $\lambda$ but only through the space-time coordinates $x^\mu(\lambda)$. Therefore, $\partial \hat{p}_{kin}^\mu / \partial \lambda = 0$ and the evolution is just given by the commutator with the Hamiltonian. 
In a similar way, we derive the equation for the time evolution of the color charge. In the Heisenberg picture, the color structure is contained in the evolution of the operators corresponding to the color matrices $\hat{T}^a$. It follows that
\begin{equation}\label{eq:HeisenbergColor}
    \begin{split}
        \frac{\diff \hat{T}^a}{\diff \lambda}  &= i[\hat{H}, \hat{T}^a] \\
        & = \frac{g}{2m} f^{abc}\left(- \mathcal{A}^b_\mu (\hat{p}^\mu_{kin} \hat{T}^c + \hat{T}^c \hat{p}^\mu_{kin}) + \mathcal{F}^b_{\mu\nu} \hat{S}^{\mu\nu} \hat{T}^c\right)\, .
    \end{split}
\end{equation}
Lastly, we consider how the particle spin changes with time. The associated Heisenberg picture operator then satisfies
\begin{equation}\label{eq:HeisenbergSpin}
    \begin{split}
        \frac{\diff \hat{S}^{\mu\nu}}{\diff \lambda} & = i [\hat{H}, \hat{S}^{\mu\nu}] = \frac{2g}{m} \hat{S}^{\rho[\mu} \hat{F}^{\nu]}_{\ \ \rho} \, ,
    \end{split}
\end{equation}
where the anti-symmetrization is denoted as $X^{[\mu\nu]}\equiv (X^{\mu\nu}-X^{\nu\mu})/2$. 

The equations of motion from \cref{eq:VelocityOperator,eq:HeisenbergKineticMomentum,eq:HeisenbergKineticMomentum,eq:HeisenbergColor,eq:HeisenbergSpin} contain all the information about the time evolution of the quantum particle in the classical medium. Together with the Heisenberg picture quantum state $|\psi \rangle_H \equiv |\psi; 0 \rangle$, they can be used to evaluate physical observables and quantum expectation values, which we denote as $\langle\hat{X}\rangle_\psi \equiv\, _H\langle \psi|\hat{X}_H|\psi\rangle_H$. The classical limit is obtained by taking the $c$-number limit of the Heisenberg equations of motion. In practice, this is done by replacing the expectation values of the quantum operators by their classical correspondents $\langle \hat{X}^\mu\rangle_\psi \rightarrow X^\mu$ with $X\in\{x^\mu, p_{kin}^\mu, T^a, S^{\mu\nu}\}$ for the coordinate, kinetic momentum, color charge and spin. When taking the classical limit, one assumes the factorization of the expectation values $\langle \hat{X}\,\hat{T}^a\rangle_\psi \rightarrow X Q^a$. Notably, the classical color charge obtained from the correspondence $\langle \hat{T}^a\rangle_\psi \rightarrow Q^a$ yields Casimir invariants which can be  different from the group theory ones $C_{2,3}$ or our previous choice from~\cref{eq:ColorChargeCasimir} with $q_{2,3}=\nc C_{2,3}$, but depend on the color state of the particle at $\lambda=0$. We can however choose an initial state where they have the values $q_{2,3}=\nc C_{2,3}$ as discussed above and in Ref. \cite{Avramescu:2023qvv}. We are interested here in the momentum broadening averaged over color using~\cref{eq:defintq}, which scales as $\langle (\delta p_i^{kin})^2\rangle_Q \propto q_2$. Thus, we  will then absorb the difference into a rescaling of $\langle (\delta p_i^{kin})^2\rangle$ by $C_2/q_2$, in order to have a classical approximation for the momentum broadening for a quark in the fundamental representation. 

Assuming the factorization of expectation values, the classical limit of \cref{eq:VelocityOperator,eq:HeisenbergKineticMomentum,eq:HeisenbergKineticMomentum,eq:HeisenbergColor,eq:HeisenbergSpin} give the set of equations of motion
\begin{subequations} \label{eq:ClassicalLimHeisenbergEq}
    \begin{align}
        &\frac{\diff x^\mu}{\diff \lambda} = \frac{p_{kin}^\mu}{m} \, , \\
        &\frac{\diff p_{kin}^\mu}{\diff \lambda} = \frac{g}{m} Q^a \left(\mathcal{F}^{\mu\nu}_a p_\nu^{kin} - \frac{1}{2} S^{\nu\rho} \,\mathcal{D}^\mu \mathcal{F}_{\mu\nu}^a\right) \, , \\
        &\frac{\diff Q^a}{\diff \lambda} = -\frac{g}{m} f^{abc} Q^c\left( \mathcal{A}^b_\mu\,  p^\mu_{kin} - \frac{1}{2} S^{\mu\nu} \mathcal{F}^b_{\mu\nu}\right) \, , \\
        &\frac{\diff S^{\mu\nu}}{\diff \lambda}=\frac{2g}{m}Q^aS^{\rho[\nu}F^{\mu],a}_{\ \ \rho} \, .
    \end{align}
\end{subequations}
One can identify the Wong's equations for the coordinate, momentum and the color charge from \cref{eq:Wong}, along with extra terms arising from the particle spin. These equations can be generalized to massless particles by using $E\diff/\diff t = m \diff/\diff\lambda$, where $t$ represents the coordinate time. By restoring the $\hbar$ factors, it turns out that the spin terms are suppressed when $\hbar \to 0$, where the classical approximation is well justified. In general, they cannot be neglected and must therefore be retained in the full calculation. However, the spin terms are also suppressed as $1/E$ (or $1/m$), and therefore do not contribute in the eikonal and static quark limits, where Wong's equations are recovered.

We emphasize that the formalism for classical colored particles with spin in classical Yang-Mills fields, derived from the quadratic form of the Dirac Hamiltonian \cite{Wong:1970fu,Heinz:1984yq} that we follow here, is not the only possibility. Alternatively, one may use the Dirac-Bergmann constrained Hamiltonian formalism to derive the covariant equations of motion for a classical massive spin-$1/2$ particle carrying non-Abelian color charge in a background Yang-Mills field \cite{Zhou:2025obm}.

\subsubsection{Kinetic and canonical momentum} \label{sec:CanBroadening}

A similar procedure can be applied to compute $\diff \hat{p}^\mu / \diff \lambda$, from which one isolates the components of the Lorentz force that contribute solely to the canonical momentum broadening. Examining the difference between the kinetic and canonical momentum given in~\cref{eq:KinCanMom} allows one to distinguish the effects of the classical medium on the particle propagation. The evolution equation for the canonical momentum may be obtained in the same way as for the kinetic momentum, see~\cref{sec:ClassicQuant}. The Heisenberg picture evolution of operators given in~\cref{eq:Ehrenfest} yields
\begin{equation} \label{eq:CanonicEqMotion}
    \begin{split}
        \frac{\diff \hat{p}^\mu}{\diff \lambda} & = i[\hat{H}, \hat{p}^\mu] \\
        & = \frac{g}{2m} \left(\hat{p}_\nu^{kin} (\partial^\mu \hat{A}^\nu) + (\partial^\mu \hat{A}^\nu) \hat{p}_\nu^{kin} - \hat{S}^{\nu\rho} \partial^\mu \hat{F}_{\nu\rho}\right)\, .
    \end{split}
\end{equation}
In the same way, the time evolution equation of the gauge field operator along the particle trajectory parametrized by $\lambda$ is given by
\begin{align} \label{eq:FieldEqMotion}
    % \begin{split}
        \frac{\diff \hat{A}^\mu}{\diff \lambda} & = i[\hat{H}, \hat{A}^\mu] \\
        &= \frac{1}{2m}\left(  \hat{p}_\nu^{kin} (\hat{D}^\nu\hat{A}^\mu)  + (\hat{D}^\nu\hat{A}^\mu)\hat{p}_\nu^{kin}+ ig \,[\hat{A}^\mu,\hat{F}_{\nu\rho}]\,\hat{S}^{\nu\rho} \right)\,\notag .
    % \end{split}
\end{align}
One may notice that the evolution of the canonical momentum and gauge field along the particle worldline from~\cref{eq:CanonicEqMotion,eq:FieldEqMotion} is consistent with the evolution of kinetic momentum from~\cref{eq:HeisenbergKineticMomentum}, since $\diff \hat{p}^\mu/\diff \lambda \, - g \diff \hat{A}^\mu/\diff \lambda = \diff \hat{p}^\mu_{kin}/\diff \lambda$ according to~\cref{eq:KinCanMom}. Using the same procedure as for the kinetic momentum, one can take the classical limit of~\cref{eq:CanonicEqMotion,eq:FieldEqMotion}, namely
\begin{subequations} \label{eq:ClassCanonicEq}
    \begin{align}
        &\frac{\diff p^\mu}{\diff \lambda} = \frac{g}{m} Q^a\left( (\partial^\mu \mathcal{A}^{\nu,a})\,p_\nu^{kin} - \frac{1}{2} S^{\nu\rho}\,( \partial^\mu \mathcal{F}^a_{\nu\rho})\right)\, , \\
        \begin{split}
        &\frac{\diff A^\mu}{\diff \lambda} = \frac{1}{m}Q^c\Big( (\partial^\nu \mathcal{A}^{\mu,c})\, p_\nu^{kin} \\
        &\phantom{\frac{\diff \mathcal{A}^\mu}{\diff \lambda}=}+ g f^{abc} \mathcal{A}^{\mu,a} \mathcal{A}^{\nu,b} p_\nu^{kin} - \frac{g}{2} f^{abc} S^{\nu\rho} \mathcal{A}^{\mu,a} \mathcal{F}^b_{\nu\rho}\Big)\, ,
        \end{split}
    \end{align}
\end{subequations}
where $\langle \hat{A}^\mu\rangle_\psi =\mathcal{A}^{\mu,a}\langle \hat{T}^a\rangle_\psi \rightarrow \mathcal{A}^{\mu,a} Q^a=A^\mu$.

For the two limiting cases, the equations of motion for the canonical momentum are as follows. For an eikonal particle propagating along the $x$-axis in the limit $E\to\infty$, the transverse components satisfy
\begin{equation} \label{eq:LorentzForceTensorCompsEik}
        \frac{\diff p^i_{eik}}{\diff t}  =  - \frac{g}{T_R} \tr{Q \left(\partial_i (\mathcal{A}_t + \mathcal{A}_x)\right)} \, ,
\end{equation}
with $i\in(y,z)$. For a static quark in the infinite-mass limit $m\to\infty$, the canonical momentum evolves according to
\begin{equation} \label{eq:LorentzForceTensorCompsStat}
        \frac{\diff p_{stat}^i}{\diff t}  =  - \frac{g}{T_R} \tr{Q (\partial_i \mathcal{A}_t)} \, .
\end{equation}
Note that in temporal gauge, the canonical momentum of a static quark does not change at all, and the kinetic momentum broadening is purely given by the contribution from the gauge field.

The expressions of the canonical momentum for eikonal quarks in~\cref{eq:LorentzForceTensorCompsEik} and static quarks in~\cref{eq:LorentzForceTensorCompsStat} are valid for any classical background field $\mathcal{A}_\mu$. Let us now specialize them to the Glasma fields introduced in \cref{sec:GlasmaFields}. The canonical momentum from~\cref{eq:LorentzForceTensorCompsEik} in the case of eikonal jets in Glasma becomes
\begin{equation}
        \frac{\diff p^{eik}_i}{\diff \tau}  =-  \frac{g}{T_R} \tr{Q f^{i}_{p,eik}} \,,
\end{equation}
where the subscript $p$ indicates the components of Lorentz force that contribute to the canonical momentum. We refer to $f_p$ as the canonical Lorentz force. The corresponding transverse components are
\begin{subequations}\label{eq:FPeik}
    \begin{align}
        &f^y_{p,eik}=\mathcal{E}^p_y-\mathcal{B}^p_z\,,\\
        &f^z_{p,eik}=\mathcal{E}^p_z+\mathcal{B}^p_y\,.
    \end{align}
\end{subequations}
These expressions resemble the Lorentz force contributing to the kinetic momentum from~\cref{eq:JetLorentzForce}. However, the canonical Lorentz force receives a contribution only from the longitudinal Glasma fields
\begin{equation}\label{eq:CanonicField} 
    \begin{split}
        \mathcal{E}^p_y &= 0 \, , \qquad \mathcal{E}^p_z = \frac{1}{\tau^2} \mathcal{A}_\eta\, , \\
        \mathcal{B}^p_y &= 0 \, , \qquad \mathcal{B}^p_z = \partial_y \mathcal{A}_x \, .
    \end{split}
\end{equation}
Here, we used the relation between the Minkowski and Milne components of the gauge field $\mathcal{A}_t=-(\sinh\eta)/\tau\,\mathcal{A}_\eta$, taken the partial derivative $\partial_z\mathcal{A}_t$ and at the end expressed the result at mid-rapidity $\eta=0$, which yields the longitudinal $z$ electric field contribution. In a similar way, the canonical broadening for static quarks becomes
\begin{equation} 
        \frac{\diff p^{stat}_i}{\diff \tau}  = - \frac{g}{T_R} \tr{Q f^{i}_{p,stat}} \, ,
\end{equation}
where the canonical Lorentz force components are determined by the electric fields from~\cref{eq:CanonicField}, namely
\begin{equation}\label{eq:FPstat}
    f^i_{p,stat}=\mathcal{E}^p_i\,.
\end{equation}

In analogy with the kinetic momentum broadening in~\cref{eq:KinMomLorentzForce}, the canonical momentum broadening can be expressed as the squared integrated Lorentz force
\begin{equation} \label{eq:CanMomLorentzForce}
        \langle (\delta p_i (\tau))^2\rangle_Q=g^2\int_0^\tau \diff \tau^\prime \int_0^\tau \diff\tau^{\prime \prime}\,\tr{ \tilde{f}_p^i(\tau^\prime)\tilde{f}_p^i(\tau^{\prime\prime})}\,,
\end{equation}
where $\tilde{f}_p^i=\mathcal{U} f_p^i\,\mathcal{U}^\dagger$ is the color-rotated canonical Lorentz force, defined analogously to \cref{eq:ColorRotLorentzForce}.
For the eikonal case, $f_p^i$ is given by \cref{eq:FPeik}, while for a static quark it is given by \cref{eq:FPstat}. 

In typical jet quenching calculations the momentum broadening of an eikonal particle is assumed to depend only on the longitudinal light-cone component of the background fields $\mathcal{A}_+ \propto \mathcal{A}_\tau +\mathcal{A}_x$ \cite{Casalderrey-Solana:2007knd}. However, we here find that for a particle propagating inside an external medium that is only true for the canonical momentum broadening $\diff p_i / \diff \tau = g \partial_i \mathcal{A}_+$, while the physical gauge invariant  kinetic momentum receives an additional contribution at the same eikonal order $g \diff \mathcal{A}_i/\diff \tau = g\, \partial_+ \mathcal{A}_i + ig^2\, [\mathcal{A}_+, \mathcal{A}_i]$, which depends on the transverse components of the background potential. Although different frameworks and kinematic regimes, the conclusion is very similar to the one found in \cite{Kar:2026vzk} for DIS di-jet production, where they claim that $\partial_i \mathcal{A}_+$ and $\partial_+ \mathcal{A}_i$ contribute with a similar significance to the field strength tensor $\mathcal{F}_{i+}$ even at leading eikonal order, and therefore $\partial_i \mathcal{A}_+$ alone is not enough to describe eikonal particle propagation in an extended medium.

\subsection{Effect of the gauge choice}
\label{sec:GaugeTransformation}

We study how gauge-dependent quantities, such as the canonical momentum broadening, are affected by the gauge condition. This provides a way to quantify the effect of gauge transformations on the background fields. We also investigate the numerical uncertainties introduced in the extraction of such gauge-dependent quantities.

\subsubsection{Gauge dependence of canonical momentum broadening}\label{subsec:gaugetransf}

In the quantum formalism, if one starts from a wavefunction given by a wave packet, the canonical momentum broadening may be interpreted as the spreading of the wave packet in momentum space, caused by the interaction with the classical background field. For the classical canonical broadening, we showed that it may be extracted from the canonical Lorentz force exerted on the particle, see \cref{eq:CanMomLorentzForce}, although its physical significance remains ambiguous. Nevertheless, because it is gauge dependent, the classical canonical broadening can serve as a diagnostic tool to quantify the effects of gauge transformations on the Glasma fields. 

Under a non-Abelian gauge transformation
\begin{equation} \label{eq:NonAbGaugeTransf}
    \mathcal{A}_\mu\rightarrow \mathcal{G}\,\mathcal{A}_\mu \,\mathcal{G}^{-1}+\frac{i}{g}\mathcal{G}\,(\partial_\mu \mathcal{G})\,,
\end{equation}
with $\mathcal{G}\in\mathrm{SU}(\nc )$, the field strength tensor transforms as 
\begin{equation}
    \mathcal{F}_{\mu\nu}\rightarrow \mathcal{G}\,\mathcal{F}_{\mu\nu}\,\mathcal{G}^{-1}\,.
\end{equation}
Consequently, the color averaged kinetic momentum broadening from \cref{eq:KinMomLorentzForce} is gauge invariant, since it is expressed as the trace of product of Lorentz force components, which are themselves given in terms of the field strength tensor from \cref{eq:ColorRotLorentzForce}. By contrast, the canonical Lorentz force for eikonal quarks [\cref{eq:FPeik}] and static quarks [\cref{eq:FPstat}] is not gauge invariant, so the resulting canonical momentum broadening given in \cref{eq:CanMomLorentzForce} is gauge dependent.     

Since the Glasma fields are boost invariant, we limit ourselves to  gauge transformations that also have the same property $\partial_\eta\,\mathcal{G}_\eta=0$, which reduces the gauge transformation from \cref{eq:NonAbGaugeTransf} for the rapidity component $\mu=\eta$ to $\mathcal{A}_\eta\rightarrow \mathcal{G}\,\mathcal{A}_\eta \,\mathcal{G}^{-1}$. As the longitudinal canonical Lorentz force for both eikonal and static quarks is determined only by the canonical electric field from~\cref{eq:CanonicField}, the gauge transformation yields
\begin{equation}\label{eq:GaugeTransffpz}
    f_p^z = \frac{1}{\tau^2} \mathcal{A}_\eta \rightarrow \frac{1}{\tau^2} \mathcal{G}\,\mathcal{A}_\eta \,\mathcal{G}^\dagger\,,
\end{equation}
Thus, although $f_p^z$ is gauge dependent, when squaring and taking the trace, the resulting canonical momentum broadening $\langle (\delta p_z)^2 \rangle_Q$ is gauge invariant. In contrast, for $\langle (\delta p_y)^2 \rangle_Q$, the inhomogenous term $\mathcal{G}\,(\partial_y \mathcal{G}^\dagger )$ generates a gauge dependence. Thus, in the boost-invariant Glasma, only the $y$ component of the canonical momentum broadening is affected by performing a gauge transformation.

\subsubsection{
Gauge field evolution along the particle trajectory
}\label{subsec:lorentzdecomp}

To study the evolution of the gauge field along the particle trajectory, we relate the change of the gauge potential 
\begin{equation} \label{eq:deltaA}
    \delta A_i(\tau)\equiv A_i(\tau)-A_i(0)
\end{equation}
to the difference between the canonical and kinetic Lorentz forces
\begin{equation} \label{eq:deltapkinpcanA}
    g\,\delta A_i^f(\tau)\equiv\delta p_i(\tau)-\delta p^{kin}_i(\tau)\,,
\end{equation}
using the classical correspondence of the kinetic momentum decomposition from~\cref{eq:KinCanMom}. These expressions are equivalent so they must satisfy $\delta A_i(\tau)=\delta A_i^f(\tau)$. The change in the gauge potential from \cref{eq:deltaA} may be expressed in terms of color components $\delta A_i(\tau)= \mathcal{A}_i^a(\tau)\,Q^a(\tau)-\mathcal{A}_i^a(0)\,Q^a(0)$. Using the relation $Q^a(\tau)=\tr{Q(\tau)\,t^a}/T_R$ along with the rotation of the color charge from \cref{eq:WongColorRotation}, the variation of the field along the trajectory becomes
\begin{equation} \label{eq:deltaAgauge}
        g\,\delta A_i(\tau)=\frac{g}{T_R}Q^a(0)\,\tr{\delta \mathcal{A}_i (\tau)\, t^a}\,,
\end{equation}
in which the change of the gauge field algebra element
\begin{equation}\label{eq:deltaArot}
    \delta \mathcal{A}_i(\tau) \equiv \tilde{\mathcal{A}}_i(\tau) - \mathcal{A}_i(0)\, 
\end{equation}
is evaluated using the parallel transported gauge field $\tilde{\mathcal{A}}_i(\tau)\equiv \mathcal{U}(\tau;0)\,\mathcal{A}_i(\tau)\,\mathcal{U}^\dagger(\tau;0)$. 

Using the expression of kinetic momentum in terms of the Lorentz force in \cref{eq:KinMomBroadLorFor} and the canonical Lorentz force introduced in \cref{sec:CanBroadening}, the change in the background gauge field along the particle trajectory can be written as the difference between the kinetic and canonical Lorentz forces
\begin{equation} \label{eq:deltaAlorentz}
        g\, \delta A_i^f(\tau)
        = \frac{g}{T_R}Q^a(0)\,\tr{\delta \mathcal{A}^f_i (\tau)\, t^a}  \, ,\end{equation}
expressed in terms of the variation of the gauge field color matrix
\begin{equation}\label{eq:deltaAforce}
    \delta \mathcal{A}_i^f(\tau) \equiv \int_0^\tau \diff\lambda \,\tilde{f}_i^A(\lambda) \, .
\end{equation}
Here, $f^A_i = f_i - f_i^p$ represent the components of the Lorentz force contributing to the gauge field variation and $\tilde{f}^A_i$ indicates that it is color rotated along the quark trajectory, as defined in \cref{eq:ColorRotLorentzForce}.

Combining~\cref{eq:deltaAgauge,eq:deltaAlorentz} yields the consistency relation
\begin{equation} \label{eq:deltaAdeltaAtilde}
    \delta \mathcal{A}_i(\tau)= \delta\mathcal{A}^f_i(\tau)\,.
\end{equation}
We now have two ways to evaluate the evolution on the transverse field along the particle trajectory, by either extracting $\delta \mathcal{A}^f_i$ which is obtained by integrating the components of the corresponding part of the Lorentz force, see~\cref{eq:deltaAforce}, or by using the color-rotated potential $\delta \mathcal{A}_i$ as in~\cref{eq:deltaArot}. Essentially, the first way corresponds to  the natural approach in the classical calculation, where one integrates the color-rotated Lorentz force $\tilde{f}_A^i$ that contributes to $\delta \mathcal{A}^f_i$ in order to solve the classical equations of motion. The second approach resembles what we expect to do in the quantum calculation, where one uses the quantum state of the particle to directly evaluate the quantum operator $\hat{A}_i$ at the time $\tau$. The two approaches are explicitly equivalent, but a reliable quantitative comparison of the classical and quantum calculations requires them to be accurate also numerically. We now turn to investigating whether this is the case in typical calculations in the Glasma fields.  

\subsubsection{Momentum broadening decomposition}\label{subsec:mombroaddecomp}

By using~\cref{eq:deltapkinpcanA}, squaring and averaging over color charge configurations as defined in \cref{eq:defintq} along with the $2$-point function given in~\cref{eq:ColorChargeIntegral}, as we did to obtain $\langle (\delta p_i^{kin})^2\rangle_Q$ in \cref{eq:KinMomLorentzForce}, one can derive a relation between the kinetic and canonical momentum broadening
\begin{equation}\label{eq:KinCanMomBroadDecomp}
    \langle (\delta p_i^{kin})^2\rangle_Q=\langle (\delta p_i)^2\rangle_Q-2g\langle \delta p_i\,\delta A_i^f\rangle_Q+g^2 \langle (\delta A_i^f)^2\rangle_Q
\end{equation}
expressed at the same $\tau$. Here, the color averaged kinetic momentum broadening is given by \cref{eq:KinMomLorentzForce} and the canonical one by \cref{eq:CanMomLorentzForce}. The remaining components can be evaluated using any of the two equivalent approaches we introduced in \cref{subsec:lorentzdecomp}. The first approach consists on integrating the corresponding components of the Lorentz force as in \cref{eq:deltaAforce}. We denote the resulting components as
\begin{subequations}\label{eq:deltaAforceSquared}
    \begin{align}
        &\langle \delta p_i \,\delta A_i^f \rangle_Q=g\int_0^\tau \diff \tau^\prime \int_0^\tau \diff\tau^{\prime \prime}\,\tr{ \tilde{f}_p^i(\tau^\prime)\tilde{f}_A^i(\tau^{\prime\prime})}\,,\\
        &\langle (\delta A_i^f)^2\rangle_Q=\int_0^\tau \diff \tau^\prime \int_0^\tau \diff\tau^{\prime \prime}\,\tr{ \tilde{f}_A^i(\tau^\prime)\tilde{f}_A^i(\tau^{\prime\prime})}\,.
    \end{align}
\end{subequations}
and replace them in \cref{eq:KinCanMomBroadDecomp}. Alternatively, by applying the consistency relation $\delta A_i^f=\delta A_i$, one may write a similar decomposition as in \cref{eq:KinCanMomBroadDecomp} for $\delta A_i^f\rightarrow \delta A_i$. The relevant terms can be obtained directly in terms of the parallel transported gauge field 
\begin{subequations} \label{eq:deltaAfieldSquared}
    \begin{align}
        &\langle \delta p_i \,\delta A_i\rangle_Q=g\int_0^\tau \diff \tau^\prime \,\tr{ \tilde{f}_p^i(\tau^\prime)\,\delta \mathcal{A}_i(\tau)}\,,\\
        &\langle (\delta A_i)^2\rangle_Q=\tr{(\delta \mathcal{A}_i)^2}\, ,
    \end{align}
\end{subequations}
with $\delta \mathcal{A}_i$ defined in \cref{eq:deltaArot}.
We therefore have two alternative ways of evaluating the decomposition of the kinetic momentum broadening, integrating different terms in the Lorentz force in  \cref{eq:deltaAforceSquared} and using the variation of gauge potentials as in \cref{eq:deltaAfieldSquared}. For the case of a jet moving in the $x$ direction in the Glasma, we consider the components $i \in (y,z)$. The transverse field $\mathcal{A}_y$ is derived from the gauge link $U_y$ via \cref{eq:GaugeField}, while the longitudinal component is $\mathcal{A}_z=\mathcal{A}_\eta/\tau$, valid at mid-rapidity $\eta=0$.

When computing the kinetic momentum broadening via the decomposition in \cref{eq:KinCanMomBroadDecomp}, the individual terms $(\delta p)^2$, $(\delta A)^2$, and $\delta p\delta A$ can be individually large, while their combination is small, leading to significant cancellations and hence larger numerical errors. Therefore, when working with gauge-dependent quantities, it is advantageous to choose a gauge in which these errors are minimized, thereby improving computational efficiency. We provide one such choice in the following subsection.

\begin{figure*}[!tbh]
\centering
\subfigure[\,Kinetic momentum broadening]{\includegraphics[width=0.9\columnwidth]
{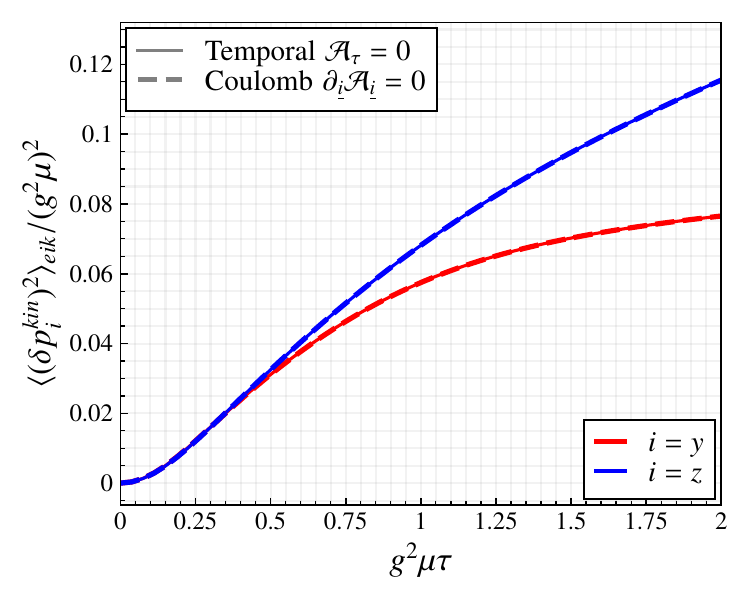}
\label{subfig:GaugeTransEffectKin}}
\quad
\subfigure[\,Canonical momentum broadening]{\includegraphics[width=0.9\columnwidth]
{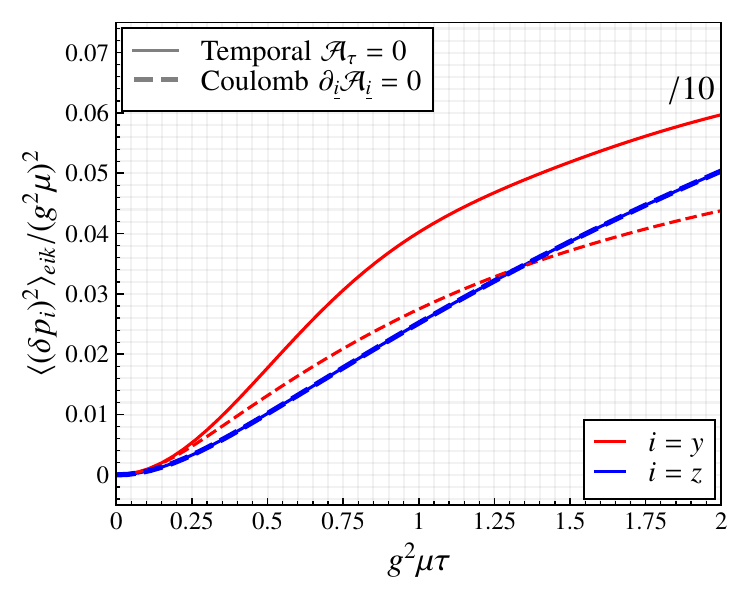}
\label{subfig:GaugeTransEffectCan}}
\caption{Kinetic \textit{(panel (a) left)} and canonical \textit{(panel (b) right)} momentum broadening for eikonal quarks, scaled by $(g^2\mu)^2$, as a function of $g^2\mu\tau$, for the $i=y$ transverse component \textit{(colored red)} and the $i=z$ longitudinal component \textit{(colored blue)}, extracted in the temporal \textit{(solid line)} and Coulomb \textit{(dashed line)} gauges. Note that the value for the $y$ component of the canonical momentum broadening in the temporal gauge is divided by $10$. 
}
\label{fig:GaugeTransEffect}
\end{figure*}

\begin{figure} [tb!] 
    \centering
    \includegraphics[width=0.5\textwidth]{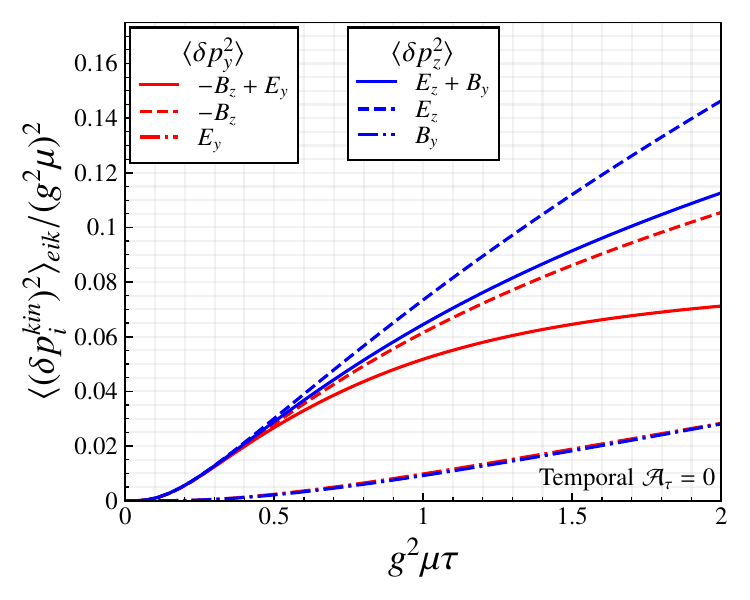}
    \caption{Kinetic momentum broadening for eikonal quarks (\textit{full line}), scaled by $(g^2\mu)^2$, as a function of $g^2\tau$, for the $i = y$ transverse component \textit{(colored red)} and the $i = z$ longitudinal component \textit{(colored blue)}, extracted in the temporal gauge $\mathcal{A}_\tau=0$. The contribution to kinetic momentum broadening from only longitudinal electric or magnetic fields \textit{(dash line)} and only from transverse fields \textit{(dash dot line)} are also represented. 
    }
    \label{fig:mom_broad_eb}
\end{figure}

\subsubsection{Coulomb gauge fixing}\label{subsec:coulombgauge}

The Glasma fields are numerically solved in the temporal gauge $\mathcal{A}_\tau=0$, but this gauge leaves a residual freedom to perform time independent gauge transformations. In order to optimize numerical efficiency in evaluating the kinetic momentum broadening via \cref{eq:KinCanMomBroadDecomp}, we further impose the transverse Coulomb gauge as an initial condition at $\tau=0$. In this gauge, the transverse components satisfy
\begin{equation}\label{eq:CoulombGauge}
\sum_{{\underline{i}}\in(x,y)}\partial_{\underline{i}}\mathcal{A}_{\underline{i}}(\tau,\vec{x}_\perp)=0\;,
\end{equation}
which can effectively reduce the magnitude of the gauge fields and therefore improve numerical stability. We only fix the Coulomb gauge condition at the initial time $\tau=0$ and allow the fields to evolve freely thereafter, so this does not constitute  a time dependent gauge transformation, since we are not imposing a gauge condition separately at different $\tau$. However, since the fields only slowly deviate from the condition~\eqref{eq:CoulombGauge}~\cite{Blaizot:2010kh,Boguslavski:2018beu}, we expect this condition to suffice for our purpose of significantly reducing the large gauge artefact component in  the field magnitude~\cite{Lappi:2003bi}. 

The above gauge condition is equivalent to the more general Coulomb gauge condition $(1/\tau^2)\,\partial_\eta \mathcal{A}_\eta+\sum_{\underline{i}}\partial_{\underline{i}}\mathcal{A}_{\underline{i}}=0$ under the assumption of boost invariance. Specifically, the transverse Coulomb gauge fixing minimizes the surface integral of the squared gauge potential
\begin{equation} \label{eq:GaugePotentialSq}
    \langle \mathcal{A}_\perp^{2}(\tau)\rangle \equiv \int \diff^2\vec{x}_\perp\, \tr{\vec{\mathcal{A}}_\perp^{\,2}(\tau,\vec{x}_\perp)}\,,
\end{equation}
where $\vec{\mathcal{A}}_\perp=(\mathcal{A}_x,\mathcal{A}_y)$. This is analogous to how the generalized Coulomb gauge fixing minimizes the volume integral of the squared gauge potential~\cite{Gubarev:2000eu,Gubarev:2000nz}. 

In the practical computation, we use the Fourier accelerated Coulomb gauge fixing, with an adaptive convergence parameter. The resulting gauge transformation $\mathcal{C}(\tau=0,\vec{x}_\perp)$ is then used to transform the Glasma fields and gauge links
\begin{subequations} \label{eq:AppliedTransformation}
    \begin{align}
        &U_{\underline{i}}(\vec{x}_\perp)\rightarrow\mathcal{C}(\vec{x}_\perp)\,U_{\underline{i}}(\vec{x}_\perp)\,\mathcal{C}^\dagger(\vec{x}_\perp+\vec{a}_{\underline{i}})\,, \\
        &\mathcal{A}_\eta(\vec{x}_\perp)\rightarrow\mathcal{C}(\vec{x}_\perp)\,\mathcal{A}_\eta(\vec{x}_\perp)\,\mathcal{C}^\dagger(\vec{x}_\perp)\,,
    \end{align}
\end{subequations}
as well as the conjugate momenta:
\begin{equation}
    \mathcal{P}^\mu(\vec{x}_\perp)\rightarrow\mathcal{C}(\vec{x}_\perp)\,\mathcal{P}^\mu(\vec{x}_\perp)\,\mathcal{C}^\dagger(\vec{x}_\perp)\,,
\end{equation} 
with $\mu \in \{\underline{i}, \eta\}$, at each lattice site $\vec{x}_\perp$ and at the initial time $\tau=0$. Further details of the numerical gauge fixing procedure are provided in \cref{app:Coulomb}. 

It is important to note that, although the transverse Coulomb gauge condition in \cref{eq:CoulombGauge} only depends on the transverse components of the background potential, the gauge transformation operator obtained from solving this condition must be applied to all the components of the background field as given by \cref{eq:AppliedTransformation}. The $\mathcal{A}_\eta$ component of the potential is therefore also modified by the imposition of the transverse Coulomb gauge condition. Remarkably, as a consequence of boost invariance, after squaring the field and taking the trace, the $(\delta A)^2$ contribution to momentum broadening is indeed gauge invariant, similarly to the case of canonical momentum broadening explained in \cref{subsec:gaugetransf}.

\begin{figure*}[!tbh]
\centering
\subfigure[\,$y$ component]{\includegraphics[width=0.9\columnwidth]
{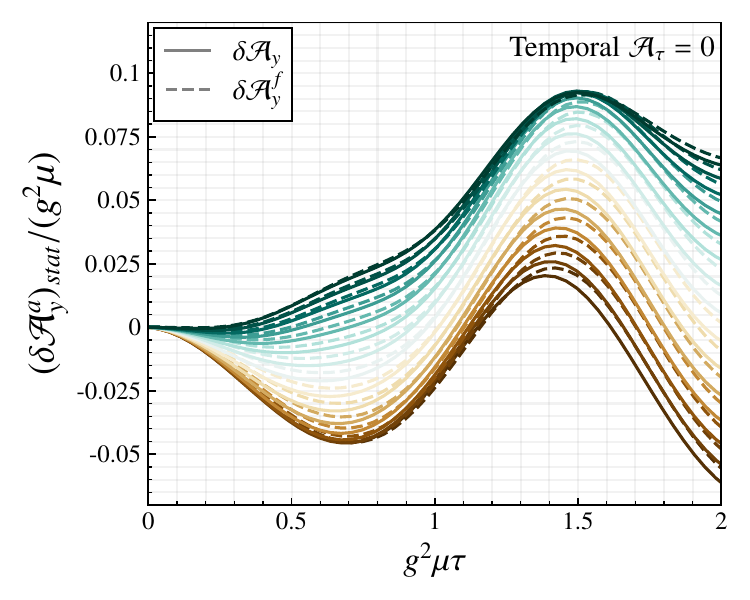}
}
\quad
\subfigure[\,$z$ component]{\includegraphics[width=0.9\columnwidth]
{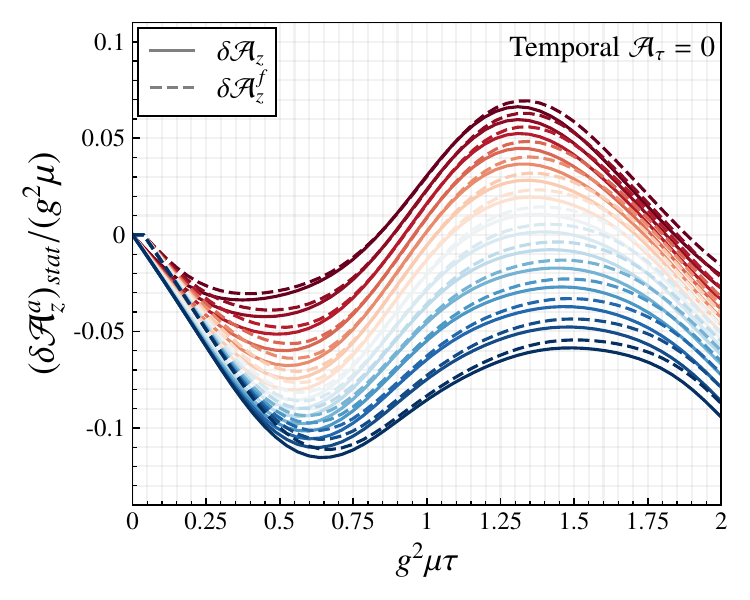}
}
\caption{
Comparison between the same color component $a$ of $\delta \mathcal{A}_{i,stat}^a(\vec{x}_\perp)$ \textit{(solid line)} and $\delta \mathcal{A}_{i,stat}^{f,a}(\vec{x}_\perp)$ \textit{(dashed line)} for static quarks, rescaled by $g^2\mu$, along the $i=y$ direction \textit{(panel (a) left)} and the $i=z$ direction \textit{(panel (b) right)}, as a function of rescaled proper time $g^2\mu\tau$. Different colors correspond to different trajectories, computed at neighboring  lattice locations $\vec{x}_\perp$, with $n_{traj}=15$ distinct trajectories shown. Results are extracted in the Glasma temporal gauge.
}
\label{fig:IntFA-static}
\end{figure*}

\begin{figure}[!tbh]
\centering
\includegraphics[width=0.9\columnwidth]
{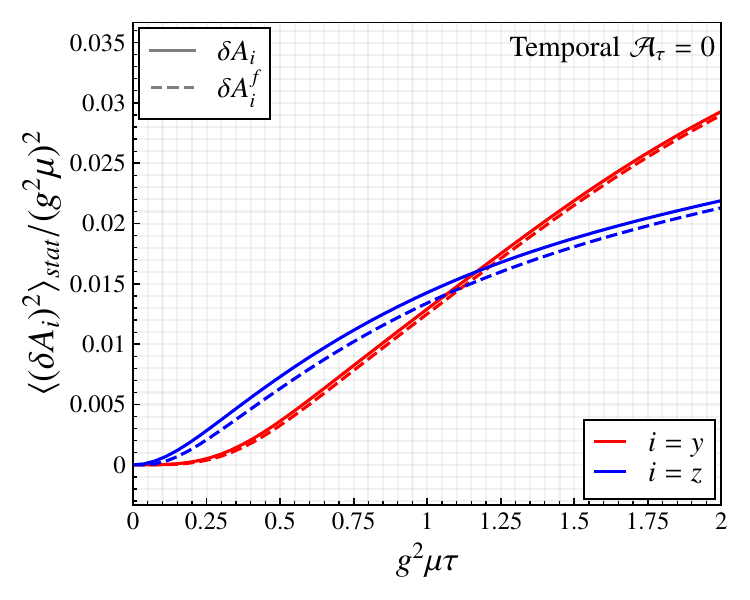}
\caption{
Comparison between the averaged $\langle(\delta A_{i})^2\rangle_{stat}$ \textit{(solid line)} and $\langle(\delta A_{i}^f)^2\rangle_{stat}$ \textit{(dashed line)} for static quarks, rescaled by $(g^2\mu)^2$, along the $i=y$ direction \textit{(colored red)} and the $i=z$ direction \textit{(colored blue)}, as a function of rescaled proper time $g^2\mu\tau$. Results are extracted in the Glasma temporal gauge.
}
\label{fig:A-AtildeComparison-static}
\end{figure}

\begin{figure*}[!tbh]
\centering
\subfigure[\,Temporal gauge]{\includegraphics[width=0.9\columnwidth]
{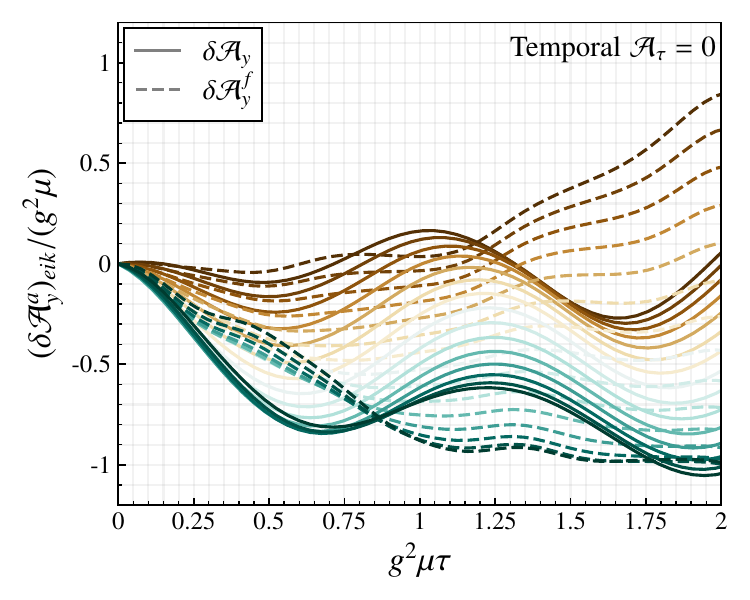}
}
\quad
\subfigure[\,Coulomb gauge]{\includegraphics[width=0.9\columnwidth]
{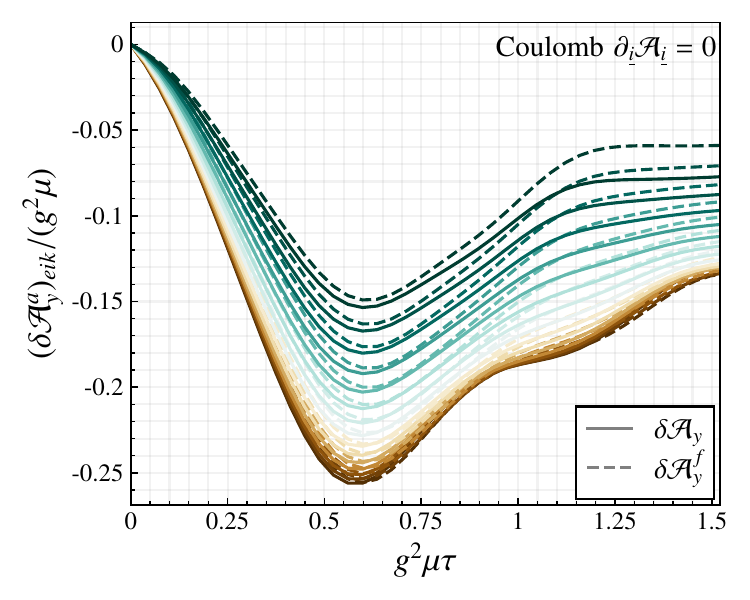}
}
\caption{Comparison between the same color component $a$ along the transverse $y$ direction of $\delta \mathcal{A}_{y,eik}^a(\vec{x}_\perp)$ \textit{(solid line)} and $\delta \mathcal{A}_{y,eik}^{f,a}(\vec{x}_\perp)$ \textit{(dashed line)} for eikonal quarks, rescaled by $g^2\mu$, in the temporal \textit{(panel (a) left)} and Coulomb \textit{(panel (b) right)} gauges, as a function of rescaled proper time $g^2\mu\tau$. Different colors correspond to different trajectories, starting from neighboring lattice locations $\vec{x}_\perp$ and computed along the quark trajectory, 
with $n_{traj}=15$ distinct trajectories shown.
}
\label{fig:IntFA-Y}
\end{figure*}

\begin{figure*}[!tbh]
\centering
\subfigure[\,Temporal gauge]{\includegraphics[width=0.9\columnwidth]
{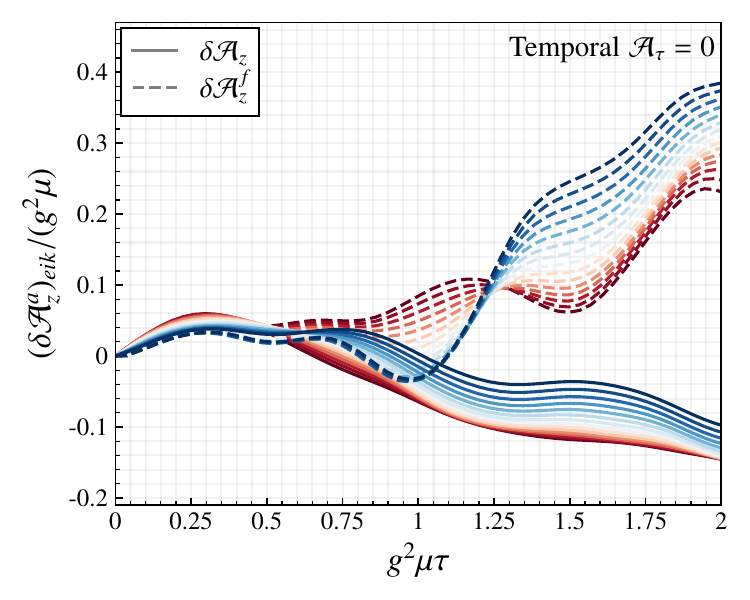}
}
\quad
\subfigure[\,Coulomb gauge]{\includegraphics[width=0.9\columnwidth]
{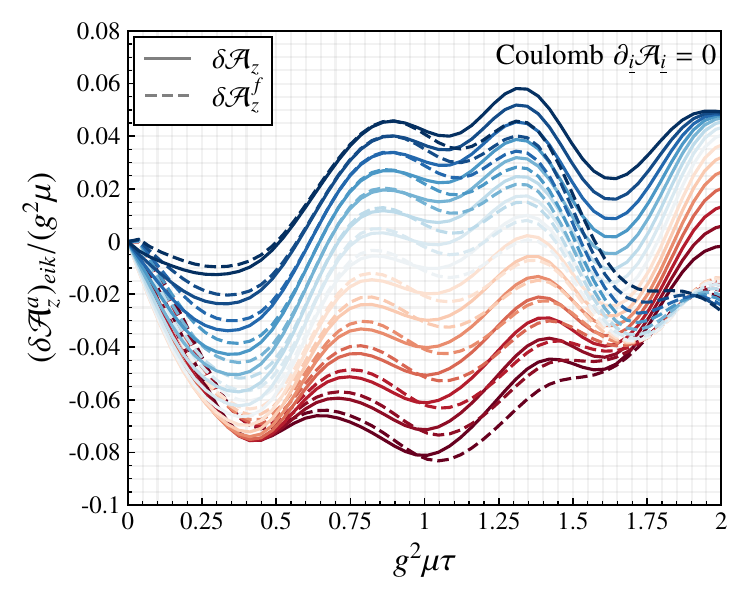}
}
\caption{Comparison between the same color component $a$ along the longitudinal $z$ direction of $\delta \mathcal{A}_{z,eik}^a(\vec{x}_\perp)$ \textit{(solid line)} and $\delta \mathcal{A}_{z,eik}^{f,a}(\vec{x}_\perp)$ \textit{(dashed line)} for eikonal quarks, rescaled by $g^2\mu$, in the temporal \textit{(panel (a) left)} and Coulomb \textit{(panel (b) right)} gauges, as a function of rescaled proper time $g^2\mu\tau$. Different colors correspond to different trajectories, starting from neighboring lattice locations $\vec{x}_\perp$ and computed along the quark trajectory, with $n_{traj}=15$ distinct trajectories shown.
}
\label{fig:IntFA-Z}
\end{figure*}

\begin{figure*}[!tbh]
\centering
\subfigure[\,Temporal gauge]{\includegraphics[width=0.9\columnwidth]
{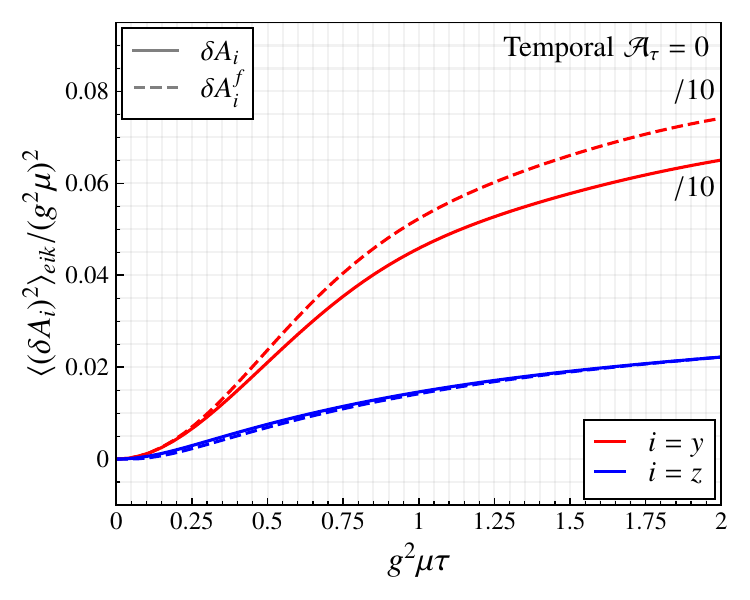}
}
\quad
\subfigure[\,Coulomb gauge]{\includegraphics[width=0.9\columnwidth]
{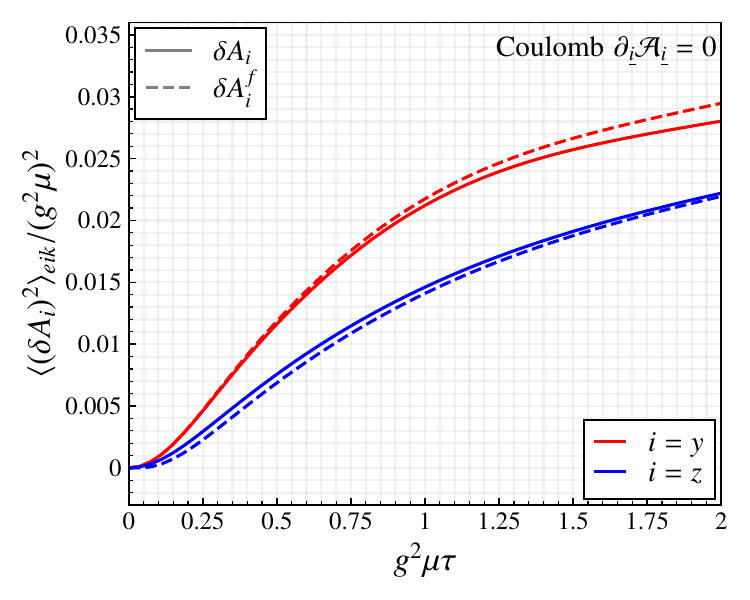}
}
\caption{
Comparison between the averaged $\langle(\delta A_{i})^2\rangle_{eik}$ \textit{(solid line)} and $\langle(\delta A^f_{i})^2\rangle_{eik}$ \textit{(dashed line)} for eikonal quarks, rescaled by $(g^2\mu)^2$, along the $i=y$ direction \textit{(colored red)} and the $i=z$ direction \textit{(colored blue)}, in the temporal \textit{(panel (a) left)} and Coulomb \textit{(panel (b) right)} gauges, as a function of rescaled proper time $g^2\mu\tau$. Note that the values corresponding to the $y$ direction in the temporal gauge are divided by $10$.
}
\label{fig:A-AtildeComparison}
\end{figure*}

\section{Results and discussion}\label{sec:results}

We will now study quantitatively how the particle dynamics are affected by performing gauge transformations on the Glasma fields, as presented in \cref{subsec:gaugetransf}. We show how the Coulomb gauge, introduced in \cref{subsec:coulombgauge}, minimizes large gauge artifacts, compared to results obtained in the Glasma temporal gauge. In the first part, we numerically evaluate the canonical momentum broadening for eikonal quarks, focusing on the $y$ component which is gauge dependent, and show how its magnitude reduces in the Coulomb gauge. We also numerically check that the kinetic momentum remains gauge invariant. In the second part, we focus on the numerical errors that accumulate when extracting gauge-dependent quantities. We use the consistency check for the accumulated gauge field introduced in \cref{subsec:lorentzdecomp} to show that numerical uncertainties are reduced in the Coulomb gauge for eikonal quarks. We also show that static quarks are not subject to this issue. In the third part, we extract the terms involved in the decomposition of the kinetic momentum broadening, as derived in \cref{subsec:mombroaddecomp}, and show that the numerical discrepancy, present in the temporal gauge, is dramatically reduced in the Coulomb gauge.

We obtain the Glasma fields by numerically solving the classical Yang-Mills equations in SU($3$), using techniques based on real-time lattice gauge theory, as described in \cref{sec:GlasmaFields}. The framework we use was developed in \cite{Muller:2019bwd} and later employed in \cite{Ipp:2020mjc,Ipp:2020nfu,Avramescu:2023qvv}. 
The results presented in this work, along with the simulation code, are publicly available\footnote{Available at \href{https://github.com/avramescudana/curraun/tree/jets}{github.com/avramescudana/curraun/tree/jets}.}. 

We simulate Glasma events with the MV model parameter from~\cref{eq:MVModel} fixed to $g^2\mu=1.5\GeV$, corresponding roughly to LHC energies. The transverse simulation region has a length $L_\perp=5\fm$ and contains $N_\perp=1024$ lattice points in each direction, yielding a lattice spacing $a_\perp=L_\perp/N_\perp$ close to the continuum limit. Moreover, $g^2\mu a_\perp\ll 1$, ensuring sufficient spatial resolution to resolve the Glasma transverse correlation domains of size $1/Q_s$ with $Q_s \approx g^2\mu$. The IR and UV regulators introduced in \cref{eq:PoissionIRUV} are set to $m_g=0.3\GeV$ and $\Lambda=10\GeV$. The lattice discretization introduces additional IR and UV cutoffs, $\lambda_{IR}=2\pi/L_\perp$ and $\lambda_{UV}=\pi/a_\perp$. Our choice of the parameters ensures that  $\lambda_{IR}<m_g$ and $\lambda_{UV}>\Lambda$, making the results insensitive to the lattice regulators. To remain in the dense Glasma regime, we also satisfy $m/(g^2\mu)\ll 1$. The resulting lattice length $g^2\mu L_\perp\approx 38$, is smaller than the typical value $g^2\mu L_\perp=100$ used in previous works~\cite{Lappi:2007ku,Ipp:2020mjc}. This choice allows us to maintain $g^2\mu a_\perp\ll 1$ for a fixed $N_\perp$, which is important when extracting the gauge field as the logarithm of a gauge link, see \cref{eq:GaugeField}, since the Taylor series used in the numerical evaluation converges better for smaller $g^2\mu a_\perp$.

The quantities we are interested in, such as kinetic or canonical momentum broadening, are averaged over the classical color charge of the quark $\langle\dots\rangle_Q$ using~\cref{eq:defintq}, which may be done analytically only for eikonal or static quarks. Further, we average over multiple Glasma events $\expconfig{\dots}$ which differ by the initial charge distribution in the MV model, see~\cref{eq:MVModel}. Lastly, unless otherwise noted, we perform averages over the lattice sites $\langle\dots\rangle_{\vec{x}_\perp}$ for quantities which are individually computed in each lattice location $\vec{x}_\perp$. For simplicity, given a classical quantity $X$, we denote its average by $\langle X\rangle$ defined as
\begin{align}
    \braket{X}\equiv 
    \expconfig{\braket{ \braket{X}_Q }_{\vec{x}_\perp} }\,.
\end{align}
We extract quantities in both the Glasma temporal gauge and the Coulomb gauge. The numerical Coulomb gauge transformation is described in~\cref{subsec:coulombgauge} and~\cref{app:Coulomb}.

\subsection{Momentum broadening in two different gauges}

We compute the color charge averaged kinetic and canonical momentum broadening for eikonal quarks from \cref{eq:KinMomLorentzForce}, and then average over Glasma events and the lattice sites where the quarks start their evolution. We denote the averaged canonical momentum broadening as $\langle (\delta p_i)^2\rangle_{eik}$ and kinetic one as $\langle (\delta p_i^{kin})^2\rangle_{eik}$ for $i\in(y,z)$. Figure~\ref{fig:GaugeTransEffect} shows the evolution of the momentum broadening in time. The subplots \cref{subfig:GaugeTransEffectKin}  and \cref{subfig:GaugeTransEffectCan} show the kinetic and  canonical momentum broadening. We use dimensionless quantities, with time multiplied by $g^2\mu$ and momentum broadening rescaled by $1/(g^2\mu)^2$ at a fixed $g^2\mu$. Since static quarks exhibit qualitatively similar behaviors, we focus only on the eikonal quarks. Both the kinetic and canonical momentum broadening are normalized by $1/\nc $ to give the expected Casimir scaling, $\langle \delta p^2\rangle\propto C_2$, as in \cite{Ipp:2020mjc,Ipp:2020nfu,Avramescu:2023qvv}.

As it should be, the kinetic momentum broadening shown in \cref{subfig:GaugeTransEffectKin} is the same in both the temporal and the Coulomb gauges. The kinetic broadening becomes larger in the $z$ than in the $y$ direction, which can be attributed to the Glasma longitudinal electric fields $\mathcal{E}_z$, which initially contribute to $\langle (\delta p_z^{kin})^2\rangle_{eik}$, to be spatially positively correlated while the longitudinal magnetic fields $\mathcal{B}_z$, which affect $\langle (\delta p_y^{kin})^2\rangle_{eik}$ at early times, exhibit anti-correlation  at some distance scales~\cite{Ipp:2020mjc}. Then in  \cref{subfig:GaugeTransEffectCan} we see that, as discussed in \cref{subsec:gaugetransf}, the canonical broadening component $\langle (\delta p_y)^2\rangle_{eik}$ is gauge dependent while $\langle (\delta p_z)^2\rangle_{eik}$ is gauge independent for the boost-invariant Glasma. One may also notice that, in the temporal gauge, the transverse $y$ component of the canonical momentum broadening is larger than the $z$ longitudinal one. As expected, the Coulomb gauge fixing reduces the magnitude of the large $y$ component by eliminating the gauge artifacts present before the gauge fixing. In this gauge, the anisotropic ordering between the $z$ and $y$ components at later times becomes the same as for the kinetic momentum broadening. 

We also studied the separate contribution of the different components of the electric and magnetic fields in \cref{eq:JetLorentzForce} to the momentum broadening of an eikonal quark. Note that each component is rotated with the same Wilson line $\mathcal{U}_{eik}$ to give a comparable result and then squared, integrated and traced as done in \cref{eq:KinMomLorentzForce} with the full Lorentz force, giving a gauge invariant quantity. Results shown in \cref{fig:mom_broad_eb} are obtained in the temporal gauge, and they match those obtained using the Coulomb gauge. As expected, initially only the longitudinal fields contribute to the kinetic momentum broadening, but the transverse fields start to give a non-vanishing contribution to the broadening already at $g^2\mu\tau \approx 0.25$. The total kinetic momentum broadening, including both the transverse and longitudinal components, is smaller in magnitude than the broadening arising only from longitudinal electric or magnetic fields. Moreover, the transverse components of the electric and magnetic fields give rise to comparable momentum broadening, while the momentum broadening coming from the longitudinal electric fields is larger than the one arising from the longitudinal magnetic fields. Intuitively, this can be traced back to the anti-correlation present in the longitudinal magnetic fields at $\tau=0$, studied in \cite{Dumitru:2013wca, Dumitru:2014nka, Ruggieri:2017ioa, Ipp:2020mjc}. This analysis emphasizes the importance of including both the longitudinal and transverse fields in studies of jet momentum broadening, and shows qualitatively that already at $g^2\mu\tau=1$ neglecting the transverse components results in an overestimation of the total kinetic momentum broadening in the Glasma fields.

\begin{figure*}[t!]
\centering
\subfigure[\,$y$ component\label{fig:DecompositionA_y}]{\includegraphics[width=0.9\columnwidth]
{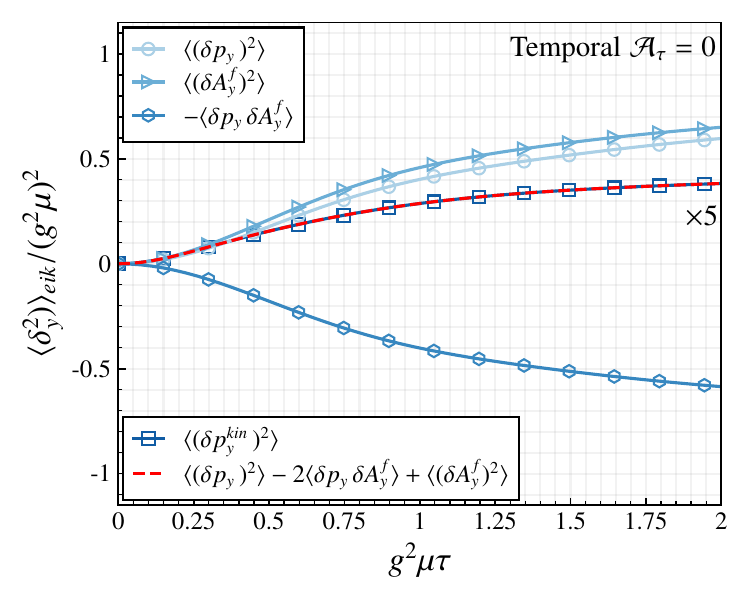}
}
\quad
\subfigure[\,$z$ component\label{fig:DecompositionA_z}]{\includegraphics[width=0.9\columnwidth]
{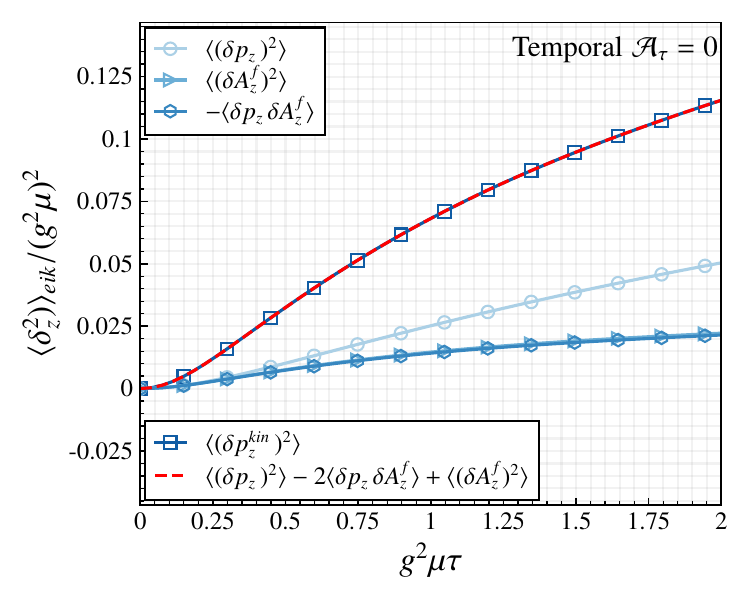}
}
\caption{
The averaged eikonal quark kinetic momentum broadening $\langle(\delta p_i^{kin})^2\rangle$ \textit{(square blue markers)} and its decomposition \textit{(red dashed line)} in the canonical momentum broadening $\langle(\delta p_i)^2\rangle$ \textit{(circle blue markers)}, the squared gauge potential variation extracted from the Lorentz force $\langle(\delta A_i)^2\rangle$ \textit{(triangle blue markers)} and the cross term $\langle\delta p_i\,\delta A_i\rangle$ \textit{(diamond blue markers)}. Each contribution is rescaled as $\langle \delta_i^2\rangle/(g^2\mu)^2$ and represented in terms of the dimensionless proper time $g^2\mu\tau$. Results are obtained in the temporal gauge, for the $i=y$ \textit{(panel (a) left)} and $i=z$ \textit{(panel (b) right)} directions. Note that both the kinetic momentum broadening and the sum of the different terms in  the decomposition along the $y$ direction are multiplied by $5$. 
    }
\label{fig:DecompositionA}
\end{figure*}

\subsection{Evolution of the gauge potential} \label{sec:PotentialEvolution}

Let us now compute the gauge field contribution to momentum broadening using the two ways described in \cref{subsec:lorentzdecomp}. This is done by either extracting $\delta \mathcal{A}_i$ as the relative difference in the parallel transported gauge field along the quark trajectory according to \cref{eq:deltaArot}, or by evaluating $\delta\mathcal{A}_i^f$ which contains the gauge field contribution to the Lorentz force, see \cref{eq:deltaAforce}. Since both calculations are physically equivalent, any discrepancy between them arises solely from the accumulation of numerical errors. 

We start by  extracting a single color component for both $\delta\mathcal{A}_i^a$ and $\delta\mathcal{A}_i^{f,\, a}$, and compare the results for $n_{traj}=15$ static or eikonal quark trajectories, starting at different lattice points. Then, we average over all the possible starting points of the quark trajectories, to compute the lattice averaged square of the gauge potential, which is then averaged over classical color and Glasma field configurations to get $\langle(\delta A_i)^2\rangle$ and $\langle(\delta A_i^{f})^2\rangle$. For eikonal quarks, we numerically extract these quantities in both the Glasma temporal gauge and after gauge transforming to the Coulomb gauge, which minimizes $\langle \mathcal{A}^{2}_\perp(\vec{x}_\perp)\rangle$ as given in \cref{eq:GaugePotentialSq}, see the discussion in \cref{subsec:coulombgauge}, and is thus expected to reduce the accumulated errors. 

For static quarks in the temporal gauge, the numerical results of the consistency check from \cref{eq:deltaAdeltaAtilde} for a single color component are shown in \cref{fig:IntFA-static} and for the color trace of the squared gauge potential, averaged over lattice locations, color charge and Glasma field configurations, in \cref{fig:A-AtildeComparison-static}. One may notice that, in the temporal gauge, the agreement is very good for both components and holds at all times $g^2\mu\tau$. For this reason, in the case of static quarks, additionally imposing the Coulomb gauge condition is not crucial, at least for this value of $g^2\mu a_\perp$. We have checked that, in this case, the imposition of the Coulomb gauge condition does not significantly modify the numerical accuracy of the calculation.

The situation is different  in the case of eikonal quarks. For both the transverse ($y$) and longitudinal ($z$) directions, the single color component of the gauge field variation along the eikonal quark trajectory, shown in \cref{fig:IntFA-Y,fig:IntFA-Z}, disagree in the temporal gauge but become compatible in the Coulomb gauge until larger values of $g^2\mu\tau$. For even smaller simulation lengths $g^2\mu L_\perp$, this agreement improves at later $g^2\mu\tau$, since in that case one is closer to the continuum limit valid for $g^2\mu a_\perp\ll 1$. 

Overall, we infer that the disagreement in the temporal gauge arises due to the accumulation of numerical errors from the color rotation. More precisely, the parallel transported gauge field $\tilde{\mathcal{A}}_i$ contains a color rotation performed with $\mathcal{U}_{eik}$ as in \cref{eq:JetColorRotation}, which picks up the gauge field component $\mathcal{A}_x$ along the trajectory of the quark moving in the $x$-direction. Therefore, the imposition of the Coulomb gauge condition not only modifies the $\mathcal{A}_{y/z}$ component, but also affects $\delta \mathcal{A}_{y/z}$ through the color rotation operator $\mathcal{U}_{eik}$. In particular, the $\mathcal{A}_x$ component of the background fields, and consequently the rate of color rotation, is largely reduced by this gauge choice, minimizing the accumulation of numerical errors and leading to a better overall agreement.

The averaged variation in the gauge field contribution squared is shown in \cref{fig:A-AtildeComparison} in both the temporal and Coulomb gauges. In the transverse direction, both the magnitude of $\langle(\delta A_y)^2\rangle$ and $\langle(\delta A^f_y)^2\rangle$, along with the disagreement between them due to numerical errors, gets significantly reduced in the Coulomb gauge. Comparing both plots in \cref{fig:A-AtildeComparison} one can see that, despite $\delta \mathcal{A}_z$ being gauge dependent, $\langle(\delta A_z)^2\rangle$ does not depend on the gauge choice. This is an expected consequence of the boost invariance of the background fields as explained in \cref{subsec:coulombgauge}.

Comparing \cref{fig:A-AtildeComparison} and \cref{fig:IntFA-static} one finds $\langle (\delta A_z)^2 \rangle$ to be similar for static and eikonal quarks, while for $\langle (\delta A_y)^2 \rangle$ they differ by a factor of approximately $20$. The physical origin of this difference can be understood from classical electrodynamics and it is especially manifest at early times, where only the longitudinal components of the Glasma field $\mathcal{E}_z$ and $\mathcal{B}_z$ contribute (see \cref{fig:mom_broad_eb}). The Lorentz force in the $z$ direction is then given by the electric field and the Lorentz force in the $y$ direction is given by the magnetic field. The electric force experienced by eikonal and static quarks will be the same as it is independent of the velocity, but while eikonal quarks are subject to a magnetic force of the same order as the electric one, static quarks are not influenced by the magnetic field at all. The same happens when we look at the canonical Lorentz force in \cref{eq:FPeik}. Therefore, if we isolate the components of the Lorentz force related to the field evolution along the particle trajectory $f_A^i = f_p^i - f^i$ we find that they will be equal for static and eikonal quarks in the $z$ direction but they can be really different in the $y$ direction. It is then not surprising that we find a similar $\langle (\delta A_z)^2 \rangle$ between static and eikonal quarks while there is around a factor of 20 difference for $\langle (\delta A_y)^2 \rangle$. The situation is more complicated at later times, where the contribution from the transverse components of the fields becomes relevant. The intuition obtained from early times is however helpful to understand how the anisotropies in the different directions arise \cite{Ipp:2020nfu}.

\begin{figure*}[t]
\centering
\subfigure[\,Temporal gauge, $y$ component]{\includegraphics[width=0.9\columnwidth]
{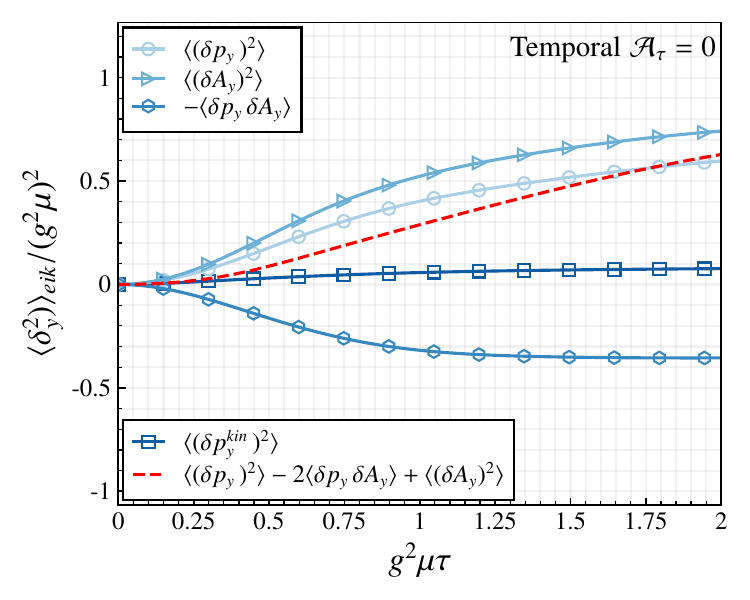}
}
\quad
\subfigure[\,Coulomb gauge, $y$ component]{\includegraphics[width=0.9\columnwidth]
{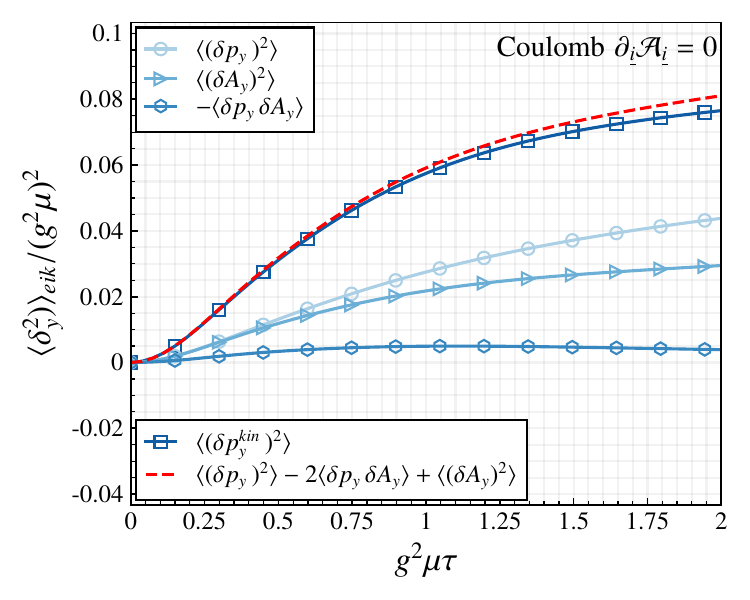}
}
\caption{
The averaged eikonal quark kinetic momentum broadening $\langle(\delta p_y^{kin})^2\rangle$ \textit{(square blue markers)} and its decomposition \textit{(red dashed line)}, the same representation as in \cref{fig:DecompositionA} but using directly the value of the gauge potential $\delta A_y$ (instead of the one obtained by integrating the corresponding component of the Lorentz force $\delta A^f_y$). Results are obtained in the temporal \textit{(panel (a) left)} and Coulomb \textit{(panel (b) right)} gauges. Note the different scale on the vertical axis in the two plots. 
    }
\label{fig:DecompoitionAtilde-Y}
\end{figure*}

\begin{figure*}[t]
\centering
\subfigure[\,Temporal gauge, $z$ component]{\includegraphics[width=0.9\columnwidth]
{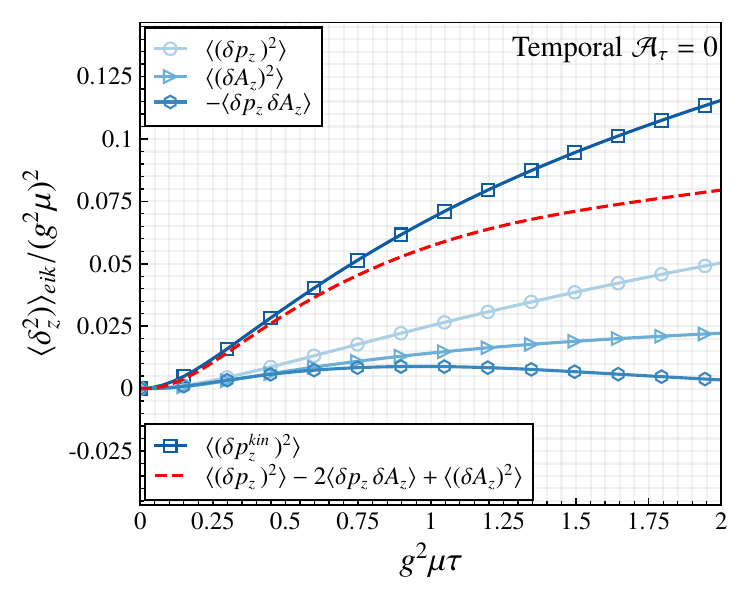}
}
\quad
\subfigure[\,Coulomb gauge, $z$ component]{\includegraphics[width=0.9\columnwidth]
{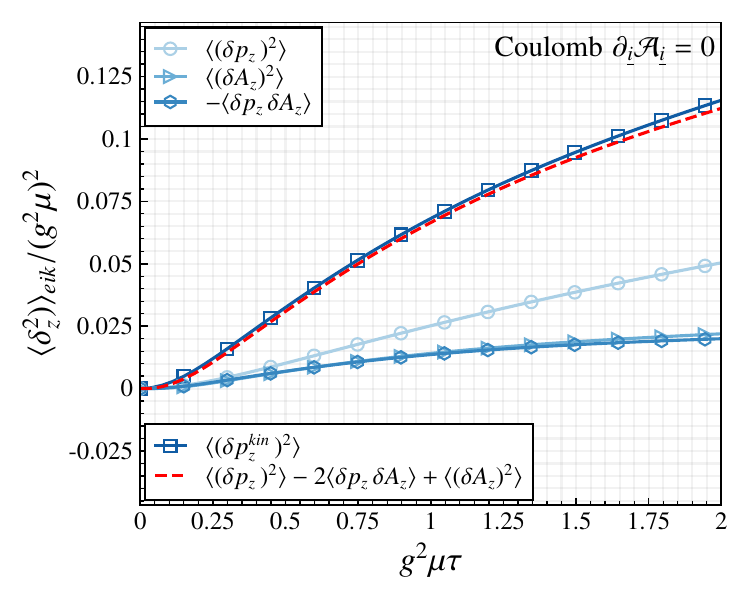}
}
\caption{
The averaged eikonal quark kinetic momentum broadening $\langle(\delta p_z^{kin})^2\rangle$ \textit{(square blue markers)} and its decomposition \textit{(red dashed line)}, the same representation as in \cref{fig:DecompositionA} but using $\delta A^f_z$ instead of $\delta A_z$. Results are obtained in the temporal \textit{(panel (a) left)} and Coulomb \textit{(panel (b) right)} gauges. 
}
\label{fig:DecompoitionAtilde-Z}
\end{figure*}

\subsection{Momentum broadening decomposition}

Having examined the consistency condition for the gauge potential in the two gauges considered, we now turn to the decomposition of the kinetic momentum broadening. In particular, we compare the kinetic momentum broadening $\braket{  (\delta p_i^{kin})^2}_Q $ evaluated directly via \cref{eq:KinMomLorentzForce} by integrating the Lorentz force, with its decomposition into three contributions: the canonical momentum square, gauge potential square, and the cross term. This decomposition follows from \cref{eq:KinCanMomBroadDecomp} and can be expressed in terms of $\delta \mathcal{A}^f$ as in \cref{eq:deltaAforceSquared} or in terms of $\delta \mathcal{A}$ as in \cref{eq:deltaAfieldSquared}. For this analysis, we focus on the eikonal quark. 

We first examine the numerical robustness of the decomposition with all terms extracted using the Glasma Lorentz forces, as given in \cref{eq:deltaAforceSquared}, in the temporal gauge. The results are shown in \cref{fig:DecompositionA}, where we use $\langle \delta _i^2\rangle$ as a generic notation for different contributions labeled in the plot legends.
For both the $y$ and $z$ components, the agreement between the direct evaluation via \cref{eq:KinMomLorentzForce} and the decomposition via \cref{eq:KinCanMomBroadDecomp} is very good, indicating strong numerical robustness when using the Lorentz-forces formulation.

For the decomposition in the $y$ direction, the three individual components are each about five times larger than the kinetic momentum broadening in magnitude, as shown in \cref{fig:DecompositionA_y}. The combined kinetic momentum broadening, which is gauge independent, therefore arises from strong cancellation among these gauge-dependent terms. Such cancellation can lead to reduced computational efficiency in the corresponding quantum calculation in the temporal gauge. This effect can be mitigated by choosing a gauge in which the field magnitude is smaller. We have checked that imposing the Coulomb gauge condition reduces their values. For the decomposition in the $z$ direction, the three individual components are approximately one order of magnitude smaller than the combined kinetic momentum broadening, as seen in \cref{fig:DecompositionA_z}. The decomposition in both $y$ and $z$ directions also holds in the Coulomb gauge, although the corresponding plots are not shown here for simplicity.

We then perform the decomposition where the gauge potential $\delta\mathcal{A}_i$ is extracted directly rather than integrating the corresponding part of the Lorentz force over time, as defined in \cref{eq:deltaAfieldSquared}. Here we perform the calculation in both the temporal and Coulomb gauges. The results are shown in \cref{fig:DecompoitionAtilde-Y} for the $y$ component and in \cref{fig:DecompoitionAtilde-Z} for the $z$ component. For both components, the agreement between the direct evaluation of the kinetic momentum via the Lorentz force \cref{eq:KinMomLorentzForce} and the decomposition using the gauge potential $\delta \mathcal{A}_i$ is poor in the temporal gauge but much better with the Coulomb gauge fixing. This behavior is consistent with, and expected from, our observations of the gauge potential in \cref{fig:IntFA-Y,fig:IntFA-Z,fig:A-AtildeComparison}, where we observed a large discretization effect in the direct extraction of the gauge potential.
Since the calculations are performed with  the same lattice spacing, the improved agreement obtained with the initial Coulomb gauge fixing indicates better numerical robustness and efficiency. Thus, we conclude that when the gauge potential is used directly in a calculation of a hard probe interacting with the Glasma, it is very beneficial to fix the Coulomb gauge condition to minimize the value of the gauge potential. This is in particular true in  the corresponding quantum formulation~\cite{Avramescu:2026qro}, where the coupling of the hard probe to the Glasma field is done via the gauge potential rather than the field strength. 

Additionally, by comparing the results in \cref{fig:DecompoitionAtilde-Y} and \cref{fig:DecompoitionAtilde-Z}, we observe that the discrepancy along $y$ is much larger than in $z$, as the larger value $\mathcal{A}_y$ with respect to $\mathcal{A}_z$ leads to more significant numerical errors that accumulate in time. In the $z$ direction, the discrepancy at intermediate to large $g^2\mu\tau$ arises not from the magnitude of the field $\mathcal{A}_z$, but from the large $\mathcal{A}_x$ field that color rotates $\mathcal{A}_z$ along the quark trajectory, as described in \cref{eq:JetColorRotation}. For the $y$ direction, both the transverse gauge fields $\mathcal{A}_y$ and the longitudinal component on the color rotation $\mathcal{A}_x$ exhibit large gauge artifacts, leading to a larger deviation due to accumulating numerical errors.

\section{Conclusion and outlook}\label{sec:summary}

We studied the propagation of a highly energetic quark through the pre-equilibrium dense gluon matter produced immediately after a heavy-ion collision, known as the Glasma. We derived the quantum evolution equations describing a particle probing a general classical non-Abelian background field and then recovered the classical equations of motion, namely Wong's equations, by taking the classical limit. In this way, we established a correspondence between the classical framework (as implemented, e.g., in Refs.~\cite{Ruggieri:2018rzi,Ipp:2020mjc,Ipp:2020nfu,Carrington:2020sww,Carrington:2021dvw,Khowal:2021zoo,Carrington:2022bnv,Avramescu:2023qvv,Avramescu:2024poa,Oliva:2024rex} and the quantum formalism that we intend to pursue in our forthcoming work~\cite{Avramescu:2026qro}, to act as a starting point for a first principles calculation of gluon radiation. 

Let us highlight two key findings. Firstly, we identified and clarified the distinction between kinetic and canonical momentum. For a particle propagating in an external gauge field, the canonical momentum, which is conjugated to the coordinates, does not always coincide with the kinetic momentum, the gauge invariant quantity that determines the momentum broadening and transport coefficients. While this distinction is irrelevant when dealing with asymptotic states in perturbation theory, it becomes important for the real-time evolution of the particle inside the medium. We derived the classical equations of motion for both canonical and kinetic momentum and used them to compute and compare the corresponding momentum broadening contributions. We found that, while in the eikonal and static quark cases the canonical momentum broadening only depends on the longitudinal components of the background field (namely $\mathcal{A}_t$ and $\mathcal{A}_x$ for a particle moving in the $x$ direction), the kinetic momentum broadening does also depend on the transverse components $\mathcal{A}_y$ and $\mathcal{A}_z$. This differs from usual jet quenching calculations, where only the $\mathcal{A}_+$ component of the background fields contributes at leading eikonal order, as they focus on measuring canonical momentum broadening far away from the interaction region. A non-trivial contribution from the transverse field components in the eikonal limit has also been recently observed in di-jet production in DIS~\cite{Kar:2026vzk}, suggesting that $\mathcal{A}_+$ may not be enough to describe the dynamics of eikonal particles propagating inside a medium.

Secondly, we examined how the numerical momentum broadening calculation is affected by the gauge choice, in particular by imposing a Coulomb gauge condition at the initial time (referred to as ``Coulomb gauge" throughout this paper). We find this gauge fixing to significantly reduce the magnitude of the gauge potential compared to calculations without it, thereby expected to improve the numerical robustness and efficiency of evaluating the kinetic momentum broadening in our forthcoming quantum calculation~\cite{Avramescu:2026qro}. 
In particular, extracting the kinetic momentum broadening involves combining different contributions--the canonical momentum broadening, the transverse gauge field squared, and the cross term between them--which may have large cancellations between them depending on the gauge. Although the final kinetic momentum is gauge invariant, these intermediate terms depend on the gauge choice, and the cancellations can become sensitive to the magnitude of the gauge potential. Choosing an appropriate gauge can therefore be important for numerical calculations. We thus recommend using Coulomb gauge fixing in similar studies, especially in quantum formulations where the gauge potential enters directly. For static and heavy quarks, however, the color rotation in the temporal gauge is trivial and numerical results converge well without imposing the Coulomb gauge condition, as shown in \cref{sec:PotentialEvolution}

This work is a first step towards developing a formalism for the quantum evolution of particles inside the Glasma phase. 
In the quantum approach, the state is represented in coordinate or canonical momentum space, and the kinetic momentum operator $\hat{p}^{kin}_i = \hat{p}_i - g\hat{A}_i$ is obtained from the separate contributions of the canonical momentum and the gauge field. In this work we analyzed these contributions within a classical framework and clarified how they combine to produce the kinetic momentum. In a forthcoming work~\cite{Avramescu:2026qro}, we plan to use classical Glasma fields in the Coulomb gauge as the background for the evolution of a quantum quark state, and use the time evolved state to evaluate physical observables such as the kinetic momentum broadening.

\section*{Acknowledgements}
We are grateful to Jo\~{a}o Barata, Yang Li, Xo\'{a}n Mayo L\'{o}pez, and David M\"{u}ller for their helpful and valuable discussions. DA and TL are supported by the Research Council of Finland, the Centre of Excellence in Quark Matter (projects 346324 and 364191). DA also acknowledges the support of the Vilho, Yrj\"{o} and Kalle V\"{a}is\"{a}l\"{a} Foundation. CL, ML, and CS are supported by European Research Council under project ERC-2018-ADG-835105 YoctoLHC; by Maria de Maeztu excellence unit grant CEX2023-001318-M and project PID2023-152762NB-I00 funded by MICIU/AEI/10.13039/501100011033; by ERDF/EU; and by Xunta de Galicia (CIGUS Network of Research Centres). CL also acknowledges support from the Ministerio de Ciencia e Innovación through the predoctoral fellowship PRE2022-102748 funded by MCIN/AEI/10.13039/501100011033. ML and CL are also supported by Xunta de Galicia under Project ED431F 2023/10.

\appendix

\section{Coulomb gauge fixing}\label{app:Coulomb}

In this Appendix, we describe the numerical implementation of the Coulomb gauge fixing introduced in \cref{subsec:coulombgauge}. We fix the residual gauge freedom from the Glasma gauge fields in the temporal gauge  $\mathcal{A}_\tau(\tau,\vec{x}_\perp)=0$ to satisfy the transverse Coulomb gauge condition $\sum_{\underline{i}}\partial_{\underline{i}}\mathcal{A}_{\underline{i}}(\tau,\vec{x}_\perp)=0$ at a fixed proper time $\tau$. It is useful to formulate the Coulomb gauge fixing as a gauge transformation procedure which minimizes the surface integral of the squared potential, namely $\langle \mathcal{A}^2_\perp(\tau)\rangle$, as defined in \cref{eq:GaugePotentialSq}. After the lattice discretization, the minimization procedure is done in terms of the functional~\cite{Davies:1987vs,Cucchieri:1995pn}
\begin{equation}\label{eq:FUG}
    \mathcal{F}_{U}[\mathcal{G}]=\dfrac{1}{8\nc  V}\sum_{\vec{x}_\perp, {\underline{i}}}\tr{U_{\underline{i}}^\mathcal{G}(\vec{x}_\perp)+U_{\underline{i}}^{\mathcal{G}\dagger}(\vec{x}_\perp)}\;,
\end{equation}
where $V=L_\perp^2$ is the transverse volume and $\nc $ is the number of colors in SU($\nc $). Here $\mathcal{G}$ is the gauge transformation and $U_{\underline{i}}^\mathcal{G}$ denotes the gauge transformed gauge links from \cref{eq:GaugeLink} namely
\begin{equation}
    U_{\underline{i}}^\mathcal{G}(\vec{x}_\perp)=\mathcal{G}(\vec{x}_\perp)\,U_{\underline{i}}(\vec{x}_\perp)\,\mathcal{G}^\dagger(\vec{x}_\perp+\vec{a}_{\underline{i}})\;.
\end{equation}
The gauge transformation is given by 
\begin{equation}\label{eq:GaugeTransf}
    \mathcal{G}(\vec{x}_\perp)=\exp\left(\mathrm{i} G(\vec{x}_\perp)\right)\;,
\end{equation}
where $G(\vec{x}_\perp)=G^a(\vec{x}_\perp)\,t^a$. The minimization of the functional~\eqref{eq:FUG} with respect to infinitesimal gauge transformations $G\to 0$ gives the lattice Coulomb gauge condition
\begin{multline}\label{eq:LatticeCoulombGauge}
    \frac{\partial \mathcal{F}}{\partial G}(G=0)\\
    =\frac{1}{16 i \nc  V}\sum_{{\underline{i}}}\left[\Delta_{-{\underline{i}}}U^\mathcal{G}_{\underline{i}}(\vec{x}_\perp)-\mathrm{h.c.}-\mathrm{trace}\right]=0\;,
\end{multline}
where we introduce the notation
\begin{equation}
    \Delta_{-{\underline{i}}}\,U_{\underline{j}}\,(\vec{x}_\perp)=U_{\underline{j}}\,(\vec{x}_\perp-\vec{a}_{\underline{i}})-U_{\underline{j}}\,(\vec{x}_\perp)\;,
\end{equation}
and the Hermitian conjugate $\mathrm{h.c.}$ is taken over the first term while the trace acts over the first two terms and contains a division by $\nc $.

The gauge fixing procedure is based on an iterative minimization of the Coulomb gauge fixing condition with respect to the gauge transformation. Since the  gauge transformation is applied locally at every lattice point $\vec{x}_\perp$, but the gauge condition involves derivatives of the links and essentially a second derivative of the gauge transformation, a local iteration procedure would converge very slowly. Instead, it is better to use Fourier acceleration following~\cite{Davies:1987vs,Berges:2013fga}  and perform updates that depend on the corresponding lattice momentum $\vec{p}_\perp\equiv (p_x,p_y)$. With the discretized derivatives, the correct concept of the square of the momentum is given by
\begin{equation}\label{eq:lattmom}
    p_\perp^2=\frac{4}{a_\perp^2}\sum_{{\underline{i}}}\sin^2\left(\frac{p_{\underline{i}} a_\perp}{2}\right)\;,
\end{equation}
which is the  eigenvalue of the discrete transverse Laplace operator $\Delta_\perp$. The maximum value is given by $p_{\perp,max}^2=8/a_\perp^2$.  The Fourier acceleration technique reduces the number of iterations needed to reach the Coulomb gauge condition and also enables low momentum modes to converge to the gauge condition in a similar time. The gauge transformation from \cref{eq:GaugeTransf} is updated according to
\begin{equation}
    \mathcal{G}^\prime(\vec{x}_\perp)=\exp\left(\mathrm{i} G(\vec{x}_\perp)\right)\mathcal{G}(\vec{x}_\perp)\;,
\end{equation}
with the gauge transformation $G(\vec{x}_\perp)$ obtained as a solution of the Poisson equation from the Coulomb gauge violation remaining after the previous transformation $\mathcal{G}(\vec{x}_\perp)$. In momentum space, this is achieved by taking the gauge transformation to be
\begin{multline}
    G(\vec{x}_\perp)=\left[F^{-1}\left(\frac{\alpha}{2}\frac{p^2_{\perp,max}}{p^2_\perp}\right)F\right]\\\sum_{{\underline{i}}}(-i)\left[\Delta_{-{\underline{i}}}U^\mathcal{G}_{\underline{i}}(\vec{x}_\perp)-\mathrm{h.c.}-\mathrm{trace}\right]
\end{multline}
where $F$ and $F^{-1}$ are the Fourier and inverse Fourier transforms, $\vec{p}_\perp$ the lattice momentum according to \cref{eq:lattmom} and $\alpha$ a parameter which controls numerical convergence, expressed in lattice units. In the non-Abelian case, this assures all momentum modes with a given $p_\perp$ converge to the Coulomb gauge condition with a comparable rate. This procedure is applied until the lattice version of the Coulomb gauge condition is satisfied. Defining the divergence of the transverse gauge field on the lattice as
\begin{equation}
    \Delta(\vec{x}_\perp)=\sum_{\underline{i}}(-i)\left[\Delta_{-{\underline{i}}}U_{\underline{i}}(\vec{x}_\perp)-\mathrm{h.c.}-\mathrm{trace}\right]\;,
\end{equation}
we see that the desired gauge condition~\eqref{eq:LatticeCoulombGauge} corresponds to $\Delta(\vec{x}_\perp)=0$. To quantify how close we are to the Coulomb gauge condition, we numerically track the trace of the squared divergence, averaged over all lattice sites
\begin{equation}\label{eq:thetacoul}
    \theta=\frac{1}{V\nc }\sum_{\vec{x}_\perp}\tr{\Delta^2(\vec{x}_\perp)}\;.
\end{equation}
In the Abelian case and in the continuum limit, convergence is reached after a single iteration with $\alpha=1/(a_\perp^2 p_{\perp,max}^2)=1/8$ using \cref{eq:lattmom}, where one should note that we are working in lattice units and  $\alpha$ is dimensionless. For the non-Abelian case, we employ an adaptive procedure to determine $\alpha^{(n)}$ in each iteration $n$. We start from $\alpha^{(0)}=0.16$, close to the Abelian estimate, then at each iteration we monitor the value of $\theta^{(n)}$ from \cref{eq:thetacoul}. If the value of $\theta^{(n)}$ oscillates, then we reduce $\alpha^{(n+1)}=\alpha^{(n)}/2$ in the next iteration. In the case of slow convergence for a few consecutive steps, we increase $\alpha^{(n+1)}=1.1\,\alpha^{(n)}$. The iterative algorithm is applied until the Coulomb gauge condition is reached with a desired numerical accuracy, and works automatically for any $N_\perp$. 

\bibliography{papers/paper}

\end{document}